\documentclass[acmsmall,screen]{acmart}

\AtBeginDocument{%
  \providecommand\BibTeX{{%
    Bib\TeX}}}

\setcopyright{acmlicensed}
\copyrightyear{2025}
\acmYear{2025}
\acmDOI{XXXXXXX.XXXXXXX}

\usepackage{graphicx}
\usepackage{listings}
\usepackage{float} 
\newfloat{lstfloat}{htbp}{lop}
\floatname{lstfloat}{Listing}
\usepackage{textcomp}
\usepackage{xspace}
\usepackage{upgreek}

\usepackage[utf8]{inputenc}
\usepackage{pgfplots}
\DeclareUnicodeCharacter{2212}{−}
\usepgfplotslibrary{groupplots,dateplot}
\usetikzlibrary{patterns,shapes.arrows}
\pgfplotsset{compat=newest}

\def\BibTeX{{\rm B\kern-.05em{\sc i\kern-.025em b}\kern-.08em
    T\kern-.1667em\lower.7ex\hbox{E}\kern-.125emX}}

\usepackage{booktabs}
\usepackage{comment}
\usepackage{cleveref}
\usepackage{ifthen}
\usepackage{enumitem}
\usepackage{soul}
\usepackage{multirow}
\usepackage{rotating}

\usepackage{makecell}
\usepackage{balance}
\usepackage{subcaption}
\usepackage{geometry}
\usepackage{fancyvrb}
\usepackage{tcolorbox}
\usepackage{arrayjobx}
\usepackage{svg}
\usepackage{algorithm}
\usepackage{algorithmic}
\usepackage{amsmath}
\usepackage{hyperref}

\usepackage{xcolor,pifont}
\newcommand*\colourcheck[1]{%
  \expandafter\newcommand\csname #1check\endcsname{\textcolor{#1}{\ding{52}}\xspace}%
}
\newcommand*\colourcross[1]{%
  \expandafter\newcommand\csname #1cross\endcsname{\textcolor{#1}{\ding{56}}\xspace}%
}

\newboolean{showcomments}
\setboolean{showcomments}{true}
\ifthenelse{\boolean{showcomments}}
{
	\definecolor{myyellow}{RGB}{255, 228, 26}
	\definecolor{myblue}{RGB}{50, 50, 220}
	\newcommand{\nb}[2]{
		{\sf
			\fcolorbox{myyellow}{yellow}{\scriptsize\textbf{#1}}%
			$\blacktriangleright$%
			{\color{myblue}\fontsize{7pt}{8pt}\selectfont\textbf{#2}}%
		}%
	}
}
{
	\newcommand{\nb}[2]{}
}

\definecolor{darkgreen}{rgb}{0,0.52,0}

\newcommand\EffBig{\textcolor{darkgreen}{$\bigstar$}}

\setlist[enumerate]{leftmargin=1.5em, label=\arabic*)}
\setlist[itemize]{leftmargin=1.5em}

\newcommand{\tool}{\textsc{Detect}\xspace} %
\newcommand{\mimicry}{\textsc{Mimicry}\xspace} %

\newcommand{\faces}{\textit{CelebA}\xspace}
\newcommand{\dogs}{\textit{Dogs}\xspace}
\newcommand{\cars}{\textit{COCO-Cars}\xspace}

\renewcommand*{\equationautorefname}{Equation}
\def\equationautorefname~#1\null{(#1)\null}

\newtheorem{definition}{Definition}

\newcommand{\head}[1]{\noindent\textbf{#1.}}
\newcommand{\changed}[1]{\textcolor{black}{#1}}

\renewcommand\footnotetextcopyrightpermission[1]{} 

\setcopyright{none}

\begin{document}

\title{Feature-Aware Test Generation for Deep Learning Models}

\author{Xingcheng Chen}
\orcid{00009-0002-0861-4093}
\email{xingcheng.chen@tum.de}
\affiliation{%
    \institution{Technical University of Munich}
    \country{Germany}
}
\email{xchen@fortiss.org}
\affiliation{%
    \institution{fortiss GmbH}
    \country{Germany}
}

\author{Oliver Wei{\ss}l}
\orcid{0009-0008-7575-0187}
\email{o.weissl@tum.de}
\affiliation{%
   \institution{Technical University of Munich}
   \country{Germany}
}
\email{weissl@fortiss.org}
\affiliation{%
  \institution{fortiss GmbH}
    \country{Germany}
}

\author{Andrea Stocco}
\orcid{0000-0001-8956-3894}
\email{andrea.stocco@tum.de}
\affiliation{%
  \institution{Technical University of Munich}
  \country{Germany}
}
\email{stocco@fortiss.org}
\affiliation{%
  \institution{fortiss GmbH}
  \country{Germany}
}

\renewcommand{\shortauthors}{Chen et al.}


\begin{abstract}
As deep learning models are widely used in software systems, test generation plays a crucial role in assessing the quality of such models before deployment. To date, the most advanced test generators rely on generative AI to synthesize inputs; however, these approaches remain limited in providing semantic insight into the causes of misbehaviours and in offering fine-grained semantic controllability over the generated inputs.
In this paper, we introduce \tool, a feature-aware test generation framework for \changed{vision-based} deep learning (DL) models that systematically generates inputs by perturbing disentangled semantic attributes within the latent space.

\tool perturbs individual latent features in a controlled way and observes how these changes affect the model's output. Through this process, it identifies which features lead to behavior shifts and uses a vision-language model for semantic attribution.
By distinguishing between task-relevant and irrelevant features, \tool applies feature-aware perturbations targeted for both generalization and robustness. 

Empirical results across image classification and detection tasks show that \tool generates high-quality test cases with fine-grained control, reveals distinct shortcut behaviors across model architectures (convolutional and transformer-based), and bugs that are not captured by accuracy metrics. Specifically, \tool outperforms a state-of-the-art test generator in decision boundary discovery and a leading spurious feature localization method in identifying robustness failures. Our findings show that fully fine-tuned convolutional models are prone to overfitting on localized cues, such as co-occurring visual traits, while weakly supervised transformers tend to rely on global features, such as environmental variances.
These findings highlight the value of interpretable and feature-aware testing in improving DL model reliability.
\end{abstract}




\maketitle

\section{Introduction}\label{sec:introduction}

Deep learning models such as deep neural networks (DNNs) and transformers have achieved impressive performance on many computer vision benchmarks~\cite{5206848}, and they are widely deployed in many software systems nowadays. However, this success often fails to translate to real-world deployments, where models encounter inputs that differ, sometimes subtly, from those seen during training~\cite{riccio2020testing, zhang2020machine}. These failures arise from two main challenges, namely \textit{limited generalization} to unseen but valid inputs, and \textit{lack of robustness} to spurious correlations that the model mistakenly relies on.
Conceptually, these problems are related to the empirical risk minimization paradigm~\cite{ERM-Vapnik1991}, which encourages DL models to optimize average performance over the training distribution, regardless of whether the features they exploit are semantically meaningful. As a result, models often learn \textit{shortcut solutions}---spurious correlations tied to background, lighting, pose, or co-occurring objects---that work on the training set but fail when exposed to real-world data distributions~\cite{Shortcut-Geirhosetal20, spurious-training-han2024}.

While deriving from the same root problem, in literature, these two challenges have been investigated separately.
While test generation techniques aim to evaluate model generalization by synthesizing new inputs that may uncover erroneous behaviors~\cite{pei2017deepxplore, zohdinasab2021deephyperion}, they often lack explainability \changed{and perturbed multiple features in an entangled manner (e.g., by injecting global Gaussian noise or simultaneously modifying lighting, texture, and object semantics). Therefore,} generated tests 
\changed{uncover failures}, but they do not clarify which feature changes caused the failure, or whether those changes are semantically relevant to the task. 
Conversely, \textit{spurious feature analysis} focuses on \textit{robustness}. They investigate whether models rely on invariant, task-irrelevant features, but typically through reactive post-hoc explanation tools such as saliency maps~\cite{gradcam-Selvaraju-2017, lrp-Bach-2015}.
\changed{This separation between input synthesis and feature diagnosis limits effective test generation and DL model improvements}.

To fill this gap, in this paper, we propose \tool, an explainable test generation framework that links \textit{test generation} and \textit{spurious feature identification} by systematically testing DNN behaviors under \textit{semantically controlled perturbations}. 
Our approach is based on the insight that both generalization and robustness concerns can be addressed through \textit{feature-aware changes} in the \changed{disentangled} latent space: by precisely manipulating \changed{individual semantic features, either relevant or irrelevant to the task}, \tool assesses not only if the DNN model generalizes to novel inputs {by varying the relevant features}, but also whether it consistently misbehaves to irrelevant ones.

To achieve this, \tool leverages a style-based generative model trained on the same distribution as the system under test (SUT), which provides access to a disentangled latent space of semantic features. By perturbing latent dimensions, we generate realistic input variants with fine-grained changes in specific attributes. This enables \textit{feature-aware test generation}, by perturbing task-relevant features to explore decision boundaries to uncover untested regions or modifying task-irrelevant features to assess prediction invariance and identify spurious dependencies. \tool incorporates a \textit{feature-aware test oracle}, which quantifies model response to controlled perturbations using a logit-based criterion. If the model's prediction significantly shifts after modifying a semantically irrelevant feature, it is flagged as a robustness failure due to spurious correlation. Conversely, changes in response to relevant features are used to test generalizability.

In our empirical evaluation on image classification and detection
tasks, \tool outperforms existing test generation and spurious feature localization methods both in revealing feature-sensitive failures, as well as robustness failures due to task-irrelevant features.
\tool reveals spurious correlations in high-accuracy models that standard accuracy metrics overlook, and distinguishes between different shortcut behaviors learned by weakly and fully supervised models. On fine-grained tasks, \tool also serves as an effective general test generator, offering flexible applicability across model types and data domains.

Our contributions are as follows:
\begin{enumerate}[topsep=0pt, partopsep=0pt, itemsep=0pt, parsep=0pt]
    \item We propose \tool, a feature-aware testing framework that perturbs semantically meaningful latent features to evaluate both generalization and robustness;
    \item We introduce a feature-aware test oracle that dually checks \textit{invariance} for irrelevant features (robustness) and explores \textit{decision boundaries} for relevant features (generalization);
    \item We show that test generation and spurious feature analysis, though distinct in goal, can be combined into a unified methodology for probing, explaining, \changed{and repairing} DNN defects.
\end{enumerate}

\section{Background}\label{sec:background}

\begin{figure*}[t]
  \includegraphics[width=0.6\linewidth]{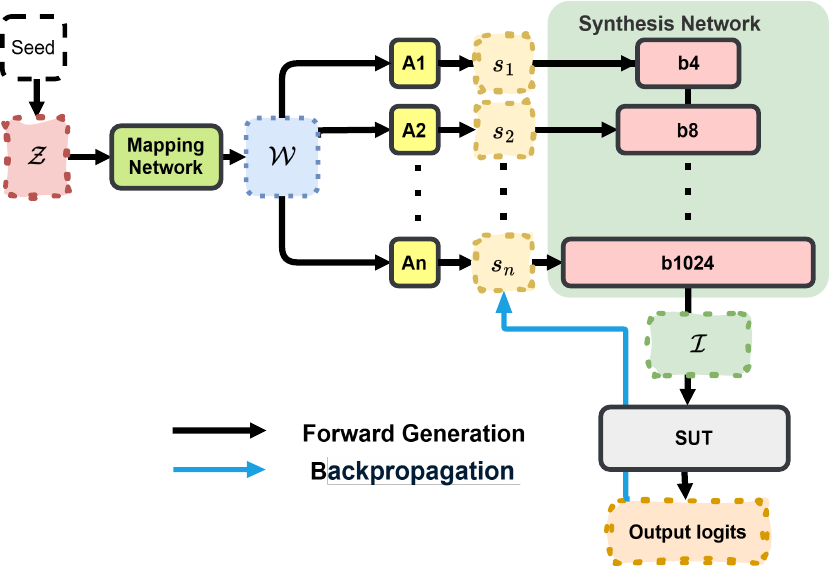}
  \caption{
        \changed{
         \textbf{System Architecture of \tool}: It comprises two main components, a StyleGAN generator and the SUT, and operates across three domains: the latent $\mathcal{S}$-space, the image space, and the output space. 
         It contains 1) a forward generation pass from an initial seed $z$ mapped into a set of style vectors $s$, which synthesize an image input passed to the SUT to obtain output logits; and 2) a backpropagation through SUT and synthesis network back to  $s$.}
}
   \label{fig:system}
\end{figure*}

In the context of deep learning testing, generative AI techniques are used due to their ability to model complex, feature-rich data distributions and synthesize new functional inputs through latent space manipulation~\cite{2023-Riccio-ICSE,2025-Maryam-ICST,dola2024cit4dnn,mimicry-weißl2025}. 
\changed{Among these, Variational Autoencoders (VAEs), Generative Adversarial Networks (GANs), and diffusion models are widely used for image generation~\cite{2025-Maryam-ICST}. However, each comes with limitations for test generation purposes. VAEs, while structurally interpretable, tend to produce low-fidelity and therefore low quality input images. Conventional GANs suffer from entangled latent spaces, making semantic control over generated features difficult. Diffusion models can generate high-quality outputs but are computationally expensive, and also lack disentanglement in their latent representations.}

Among these, recent approaches based on style-based GANs (StyleGAN)~\cite{stylegan1-Karras2021,stylegan2-karras2020,stylegan3-Karras2021}, offer high controllability due to disentangled and interpretable latent spaces. These models enable precise control over high-level image characteristics, referred to as \textit{features}, such as texture, pose, lighting, or background, through separate style and content layers~\cite{mimicry-weißl2025}. This controllability makes StyleGAN particularly well-suited for feature-aware test generation, where the goal is to manipulate specific semantic attributes of an image. As such, the remainder of this section briefly details the StyleGAN architecture.

StyleGAN~\cite{stylegan1-Karras2021,stylegan2-karras2020,stylegan3-Karras2021} comprises two main components (see top left of \autoref{fig:system}): a mapping network and a synthesis network. The generation process begins with a randomly sampled latent code $z \in \mathcal{Z}$, which the mapping network then transforms into an intermediate latent space $\mathcal{W}$, that better captures disentangled image features. Subsequently, the synthesis network applies learned affine transformations to produce per-layer style vectors $s$, resulting in the StyleSpace ($\mathcal{S}$). These style vectors modulate the convolutional activations at each block, allowing the generator to control image features at different semantic and spatial levels—early (coarse) layers influence global properties (e.g., pose, shape), while later (fine) layers affect localized details (e.g., texture, lighting).
Each style vector in $\mathcal{S}$ consists of multiple \textit{channels}, where each channel controls a distinct and semantically meaningful visual attribute, thereby enabling precise and localized image manipulations. 
This property allows us to trace, perturb, and interpret the influence of individual style channels on downstream predictions.

\begin{figure}[t]
\centering
\includegraphics[width=0.6\linewidth]{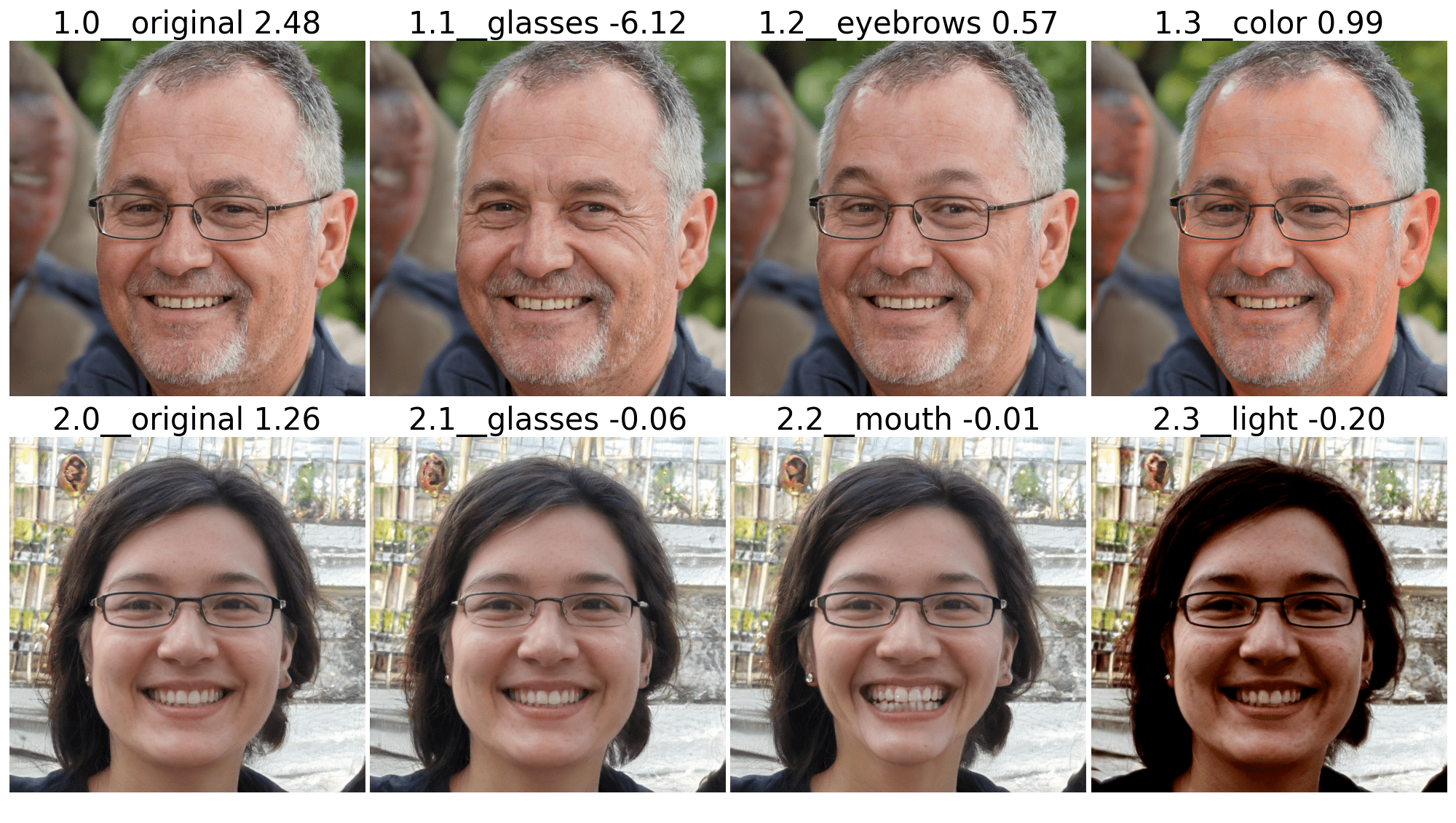}
\caption{Example perturbations in StyleSpace for glasses classification.
Each row shows a base image (left) and variants generated by perturbing a single feature (i.e., glasses, eyebrows, skin color). Titles: channel index, feature label, model output logit.
}
\label{fig:manual-explore}
\end{figure}

\autoref{fig:manual-explore} shows several examples related to the glasses detection task. By perturbing task-relevant features, one can control the presence or absence of glasses, resulting in a true negative prediction by the classifier on image 1.1 (where glasses are absent), but also a false negative prediction on image 2.1 (where instead glasses are present).
Modifications on task-irrelevant features--such as eyebrows, facial expression, or sunlight--cause a drop in classifier confidence (such as images 1.2 and 1.3), and in some cases, even result in misclassifications (such as images 2.2 and 2.3). These findings highlight a critical challenge: not all influential features are task-relevant.

\section{Feature-Aware Test Oracle}
\label{sec:feature-aware-strategy}

A central component of our testing framework is a \textit{feature-aware test oracle} that defines behavioral expectations based on the semantic relevance of input features. Unlike traditional test oracles that apply a uniform correctness criterion (e.g., \textit{misclassification}) across all inputs, our oracle distinguishes between \textit{task-relevant} and \textit{irrelevant} features, assigning distinct testing objectives to each category. \changed{First, we define a feature as a semantically interpretable factor in the latent space, whose controlled manipulation leads to a measurable change in the generated input.} 

\head{Formalization}
Let $\mathcal{T}$ be a $K$-class classification task, and let $\mathbf{F}(x) \in \mathbb{R}^K$ be the model's output logits for input $x$, with $\mathbf{F}_t(x)$ representing the logit for the target class $t \in \{1, \dots, K\}$. Let $\phi \in \mathcal{F}$ be a controllable semantic feature of the input, manipulated through a generative model into a new input $x'$.

\begin{definition}[Task-Relevant Feature]
A feature $\phi$ is \textbf{task-relevant} with respect to task $\mathcal{T}$ (denoted $\phi \in \mathcal{F}_{\text{rel}}^\mathcal{T}$) if perturbing $\phi$ induces a meaningful change in the ground truth label or alters the semantics necessary for correct prediction under $\mathcal{T}$.
Let $x \sim_\phi x'$ denote that $x$ and $x'$ differ only in feature $\phi$. Then, a feature $\phi$ is task-relevant if:
\begin{equation}
    \exists\, (x, x') \text{ such that } x \sim_\phi x', \quad \text{label}(x) \neq \text{label}(x').
\end{equation}
\end{definition}

\begin{definition}[Task-Irrelevant Feature]
A feature $\phi$ is \textbf{task-irrelevant} (denoted $\phi \in \mathcal{F}_{\text{irr}}^\mathcal{T}$) if changes in $\phi$ do not alter the ground true label under task $\mathcal{T}$, and are semantically non-essential to the prediction objective. Formally,
\begin{equation}
    \forall\, (x, x') \text{ such that } x \sim_\phi x',\quad \text{label}(x) = \text{label}(x').
\end{equation}
\end{definition}

\begin{definition}[Spurious Feature]
A feature $\phi \in \mathcal{F}_{\text{irr}}^\mathcal{T}$ is \textbf{spurious} if there exists a pair of inputs that differ only in $\phi$, and this change exercises a significant influence on the model's prediction. Formally, $\phi$ is spurious if:
\begin{equation}
    \exists\, (x, x') \text{ such that } x \sim_\phi x', \quad \Delta_{\text{logit}}(\phi) = \left| \mathbf{F}_t(x') - \mathbf{F}_t(x) \right| > \tau, 
    \label{eq: conf}
\end{equation}
where $\tau$ is a predefined confidence threshold.
\end{definition}

Our oracle supports two complementary testing goals:

\head{Oracle misclassification}
It applies to relevant features for behaviour exploration (i.e., DNN generalization).
For $\phi \in \mathcal{F}_{\text{rel}}^\mathcal{T}$, we perturb inputs to explore semantic behaviours. Prediction changes here are expected and indicative of the model's sensitivity to meaningful feature variation. Misclassifications near these transitions signal overfitting or under-generalization.

\head{Oracle confidence} 
It applies to irrelevant features for DNN robustness assessment.
For $\phi \in \mathcal{F}_{\text{irr}}^\mathcal{T}$, the model should behave invariantly. Thus, this oracle monitors $\Delta_{\text{logit}}(\phi)$ as a proxy for overreliance on non-causal features. Large shifts indicate spurious behavior.

In practice, due to the hierarchical and interdependent nature of some features, the latent space of the generator cannot be perfectly disentangled. We select $\mathcal{S}$ space as the manipulation space, as it achieves a disentanglement score of 0.75, much higher than 0.54 measured from $\mathcal{W}$ space \cite{sspace-Wu2021}. In this sense, $\mathcal{S}$ space represents the most disentangled and controllable space currently attainable for our purpose. 
To ensure methodological rigor, we adopt a conservative attribution rule: when a feature modification introduces a clear relevant semantic change but also causes small incidental adjustments, we treat the feature as relevant and ignore such minor co-effects. For example, modifying the presence of glasses while slightly affecting nearby shading or eyebrow shape.
\section{Approach}\label{sec:approach}

\begin{figure}[t]
  \includegraphics[width=1\linewidth]{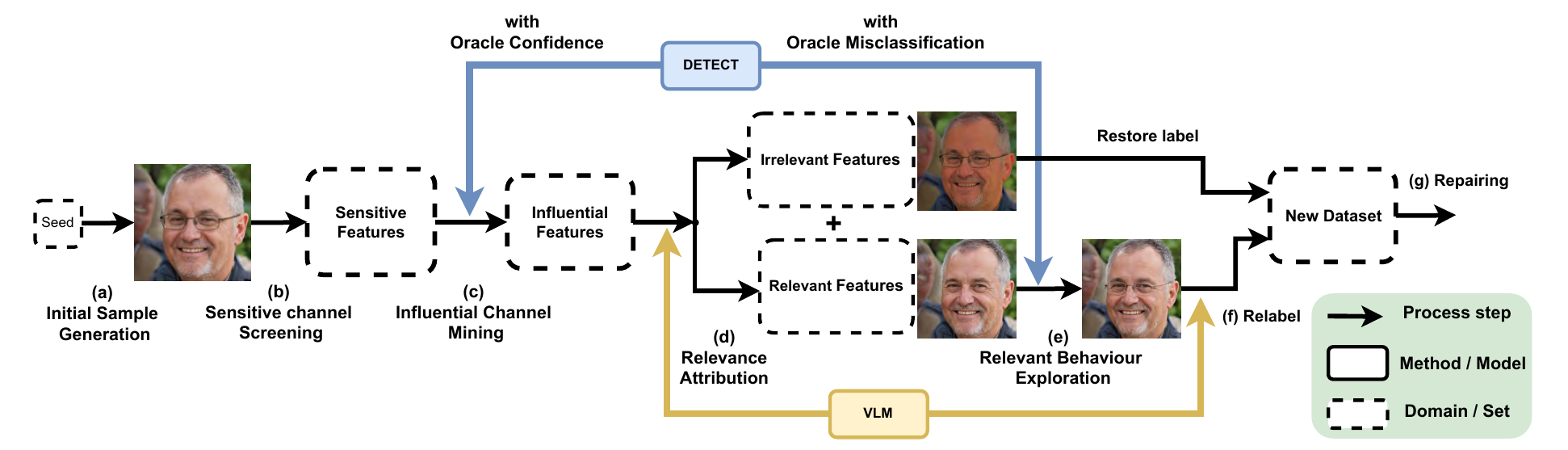}
  \caption{Sequential Pipeline of \tool. 
}
   \label{fig:pipeline}
\end{figure}

\autoref{fig:system} illustrates the workflow of our proposed framework, which we call \tool. 
\tool first samples a random latent seed, which is passed through a StyleGAN trained on the same domain as the SUT to generate an image (step Initial Sample Generation). 
To understand which channels in each latent feature $s_i$ affect the model's prediction, \tool estimates the sensitivity of each channel on the target logit with explainable AI (XAI) methods (step Sensitive Channel Screening). 
Then, \tool invokes an oracle-aware perturbation loop twice with different oracles. In the first iteration (step Influential Channel Mining), \tool perturbs all sensitive \changed{features} along the direction that reduces the SUT's confidence.
\changed{and collect the \textit{influential} features that cause significant output changes in the model's output.}
\tool then identifies whether these are task-relevant \changed{using a Vision Language Model (VLM)} \changed{(step Relevant Attribution)}. If a feature is found to be semantically irrelevant yet influences the output, \tool labels it as \textit{spurious}. On the other hand, for features deemed relevant, \tool performs a second perturbation guided by the Oracle Misclassification to explore the SUT behaviours \changed{(step Relevant Behavior Exploration)}. The second perturbation is optimized via hill climbing until decision-boundary tests are obtained. These new inputs are then relabeled by the VLM, producing a test suite that captures both relevant and spurious feature variations.

In the remainder of the section, we describe each step in greater detail.

\subsection{Initial Sample Generation}\label{sec:initial}

First, a random latent seed $z$ is generated and passed through a StyleGAN mapping network 
to obtain an intermediate latent vector $w \in \mathcal{W}$ (\changed{\autoref{fig:system}} ).
This vector is then transformed into a set of style vectors $s_i$, each of which modulates a specific synthesis block and spatial resolution during image generation.
Then, the generated image $\mathcal{I}$ is fed into the SUT to obtain class logits. \tool uses the logit output of the target class because it provides a direct, unnormalized measure of the SUT's confidence, allowing precise assessment of the model's sensitivity to controlled changes in specific latent features.

\subsection{Sensitivity Screening Methods}\label{sec:sensitivity}

 \changed{Unlike most search-based latent space test generation methods exploring the entire high-dimensional feature space~\cite{mimicry-weißl2025, 2025-Maryam-ICST} (typically spanning thousands of latent dimensions), \tool employs XAI-based sensitivity estimation to efficiently screen sensitive features and directions, reducing the subsequent perturbation search space.}

We define the sensitivity of the SUT's target logit $ y[t] $ to the $i$-th style vector $s_i$, where $\mathbf{G}(s)$ denotes the synthesis network that synthesizes the image from the style vector $ s = \{s_1, ..., s_n\}$ and  $s_i \in \mathbb{R}^d$ (the length of style vector $d$ varies among different layers), i.e., 
\begin{equation}\label{eq:sens}
    \alpha_i^{(t)} = \mathbf{S}\left(y[t], s_i\right) = \mathbf{S}\left((\mathbf{F}_t\circ \mathbf{G})(s), s_i\right ) \in \mathbb{R}^d,
\end{equation}

where $y[t]$ denotes the logits of SUT for the target class $t$ and the sensitivity $\alpha_i$  that has the same dimension as $s_i$. 

In this work, we adopt three established XAI methods, applied not to the image space, but to the disentangled latent space.

\head{Gradient Saliency~\cite{saliency_simonyan2013deep}} It computes the sensitivity of the target logit $y[t]$ with respect to each style vector $s_i$ by backpropagation through the composed function $\mathbf{F}_t\circ \mathbf{G}$, i.e.,

\begin{equation}\label{eq:backpr}
    \alpha_i^{(t)} = \frac{\partial y[t]}{\partial s_i} = \frac{\partial (\mathbf{F}_t\circ \mathbf{G})(s)}{\partial s_i}.
\end{equation}
This serves as a first-order approximation of the local sensitivity of the output to each style component.

\head{SmoothGrad~\cite{smoothgrad-smilkov2017}} 
While pure gradients are often noisy and sensitive to local variations in the input space, we adopt SmoothGrad to improve stability by averaging gradients over noise-perturbed input. We sample $N$ noisy  $\{s^{(j)} = s + \epsilon^{(j)}\}_{j=1}^N$ with $\epsilon^{(j)} \sim \mathcal{N}(0, \sigma^2 I_d)$, and then estimate for the sensitivity of channel $s_i$ to output logit $y[t]$ as: 

\begin{equation}\label{eq:smooth}
    \alpha_i^{(t)} = \frac{1}{N} \sum_{j=1}^N \frac{\partial y[t]}{\partial s_i^{(j)}} 
    = \frac{1}{N} \sum_{j=1}^N \frac{\partial (\mathbf{F}_t\circ \mathbf{G})(s^{(j)})}{\partial s_i^{(j)}}.
\end{equation}

\head{Finite-Difference Approximation~\cite{fda}}
This quantifies the local effect of each latent channel via output variation 
by computing a gradient-free sensitivity approximation based on finite differences. This method is applicable even when the composed model $\mathbf{F} \circ \mathbf{G}$ is non-differentiable or analytically inaccessible.
For each channel $s_{i,c}$ of style vector $s_i$, we apply a small perturbation $\Delta$ and the sensitivity of channel and estimate sensitivity as:

\begin{equation}\label{eq:fda}
    \alpha_{i,c}^{(t)} = \frac{ \mathbf{F}_t \circ \mathbf{G}(s_i + \Delta \cdot e_c) - \mathbf{F}_t \circ \mathbf{G}(s)}{ \Delta },
\end{equation}
where $e_c$ is the $c$-th standard basis vector in $\mathbb{R}^d$.

After computing the sensitivity scores, \tool retrieves a channel-wise attribution map for each layer in the synthesis network. Since each style vector can have up to a thousand latent channels, only the top-$k$ channels with the highest absolute sensitivity scores are retained for further analysis. 
Channels outside this set are considered inactive or behaviorally negligible with respect to the SUT.
While we use absolute sensitivity for ranking, we preserve the sign of each score. In fact, the directionality of the attribution (i.e., whether increasing/decreasing a channel pushes the output logit up or down) is used by the perturbation procedure to apply signed directional changes to the most influential channels.

\subsection{Perturbation Loop}\label{sec:Perturbation}

\begin{figure*}[t]
  \includegraphics[width=0.8\linewidth]{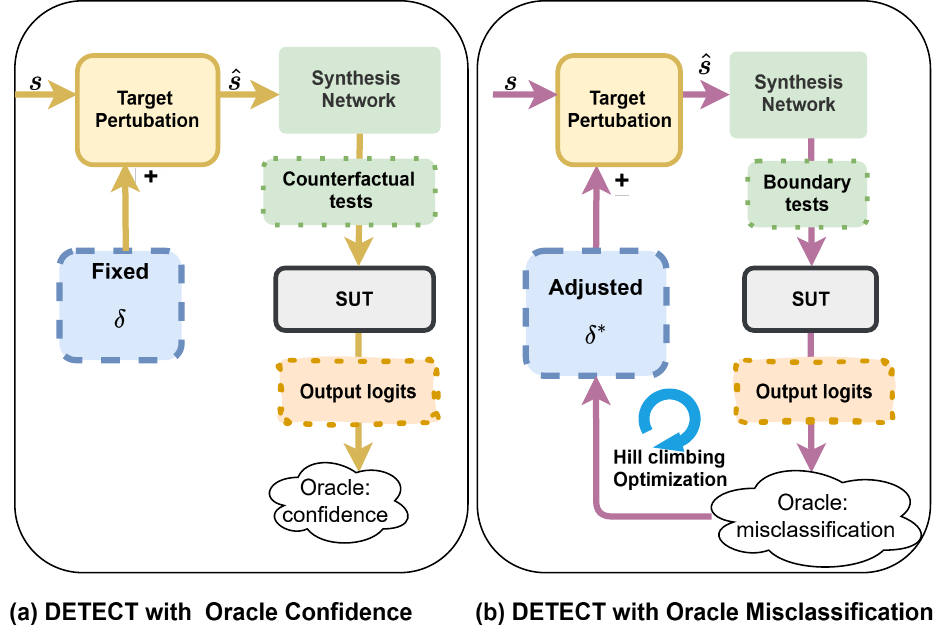}
  \caption{\tool with Oracle Confidence and Misclassification.
  }
   \label{fig:perturbation}
\end{figure*}

Once sensitive channels have been identified, \tool invokes oracle-specified perturbations to evaluate their behavioral influence, as outlined in stages (c)~Influential Channel Mining and (e)~Relevant Behaviour Exploration in \autoref{fig:pipeline}.
This two-stage perturbation is guided by the feature-aware oracles described in \autoref{sec:feature-aware-strategy}. \changed{
\autoref{fig:perturbation} provides an overview of these two configurations.
In the first configuration (a), the perturbation $\delta$ is fixed, and the Oracle Confidence guides the identification of influential features by measuring the change in output logits.
In the second configuration (b), the perturbation $\delta$ is iteratively adjusted through hill-climbing optimization, using misclassification as the oracle, to generate tests located near the decision boundary.
The detailed algorithmic steps of this loop are provided in \autoref{alg:perturbation}.}

\autoref{alg:perturbation} illustrates the test generation by perturbing latent channels under a specified oracle. Here, $\mathcal{C}$ denotes a set of selected channels (identified via sensitivity analysis or semantic relevance), and $\mathcal{O}$ specifies the behavioral oracle, either confidence or misclassification.
\begin{algorithm}[t]
\caption{\textsc{ChannelPerturb}$(s, \mathbf{F}, \mathcal{C}, \mathcal{O}, \tau)$}
\label{alg:perturbation}
\begin{algorithmic}[1]
\REQUIRE Style vector $s$, classifier $\mathbf{F}$, channel set $\mathcal{C}$, oracle $\mathcal{O}$, threshold $\tau$
\STATE Initialize perturbed result set $\mathcal{R} \leftarrow \emptyset$
\FOR{each $(l, c)$ in $\mathcal{C}$}
  \STATE Perturb $\hat{s}_l[c] = s_l[c] + \delta$
  \STATE Generate $\hat{x} = \mathbf{G}(\hat{s})$, $\hat{y} = \mathbf{F}(\hat{x})$
  \IF{$\mathcal{O} = \texttt{confidence}$ and $\Delta_{confidence} > \tau$}
    \STATE Add $(l, c, \hat{x})$ to $\mathcal{R}$
  \ELSIF{$\mathcal{O} = \texttt{misclassification}$ and $\mathcal{L}(\hat{y})\neq \mathcal{L}(y)$}
    \STATE Refine $\delta^*$ via hill climbing to decision boundary
    \STATE Add $(l, c, \hat{x}^*)$ to $\mathcal{R}$
  \ENDIF
\ENDFOR
\STATE \textbf{return} $\mathcal{R}$
\end{algorithmic}
\end{algorithm}

The main loop of the algorithm enumerates each channel $(l, c) \in \mathcal{C}$, and then applies a perturbation $\delta$ on each selected channel $s_{l,c}$ scaled by a fixed step size $\epsilon$ (Line 3). The step size $\epsilon$ is chosen to induce a visible semantic change in the generated image, while avoiding artifacts or unrealistic distortions~\cite{sspace-Wu2021,2025-Maryam-ICST}. 
\changed{The step size $\epsilon$ is determined by the scale of the latent space, independent from the dataset or the downstream SUT, since the values in $\mathcal{S}$ space after affine transforms lie in stable numeric ranges~\cite{stylegan1-Karras2021}.}
The perturbation step $\delta$ is constructed to reduce the model's confidence in the current prediction. 
Its formulation differs slightly between binary and multiclass classification, as shown below:

\begin{equation}\label{eq:delta}
\delta = \begin{cases}
-\epsilon \cdot \text{sign}(\alpha_{i,c}^{(t)}) \cdot \text{sign}(y[t]), & \text{binary,} \\
-\epsilon \cdot \text{sign}(\alpha_{i,c}^{(t)}), & \text{multiclass.}
\end{cases}
\end{equation}

This design ensures that the perturbation moves in the direction opposite to the gradient-derived influence, optionally adjusting for logit polarity in binary classification. 

The perturbed latent $\hat{s}$ is then used for synthesizing the perturbed image $\hat{x} = \mathbf{G}(\hat{s})$, and obtain the system's response $\hat{y} = \mathbf{F}(\hat{x})$ (Line 4). The logit $\hat{y}$ is analyzed according to the selected oracle. 
Under the confidence-based oracle, \tool records the perturbed image if the drop of confidence in the target logit exceeds a predefined threshold, indicating SUT's prediction relies on the perturbed feature (Lines 5-6).
Under the misclassification-based oracle, \tool checks whether the predicted label changes (Lines 7-9). We assume $\mathcal{L}(\cdot)$ returns the predicted class. For binary classification, $\mathcal{L}(\cdot)$ checks a sign in the single logit; for multiclass, it takes the top-1 class. If misclassification occurs, \tool refines the perturbation $\delta$ using a hill-climbing strategy to find the minimal modification required to cross the decision boundary, thereby producing more precise boundary examples $\hat{x}^*$.

\autoref{alg:perturbation} is executed twice: first, to probe all sensitive channels using the confidence-drop oracle; and second, only on task-relevant candidates to explore the model's behavior boundaries using the misclassification oracle.
Specifically, \tool first obtains channel-wise sensitivities using one of the methods in \autoref{sec:sensitivity} and then extracts the top-$k$ most sensitive channels $c$ by sorting their absolute sensitivity scores to form a candidate set $\mathcal{C}_{sensitivity}$. 
Then we apply the first-stage \textsc{ChannelPerturb} with the confidence-based oracle to isolate latent channels whose perturbation causes significant behavioral shifts. These are referred to as \textit{influential} channels. 
However, at this point, \tool does not yet distinguish whether the altered features are task-relevant or irrelevant, which is done in the last step.

\subsection{Relevance Attribution, Relabel, and Repair}\label{sec:relevance-attribution}

\begin{figure}[t]
\centering
\includegraphics[width=0.6\linewidth]{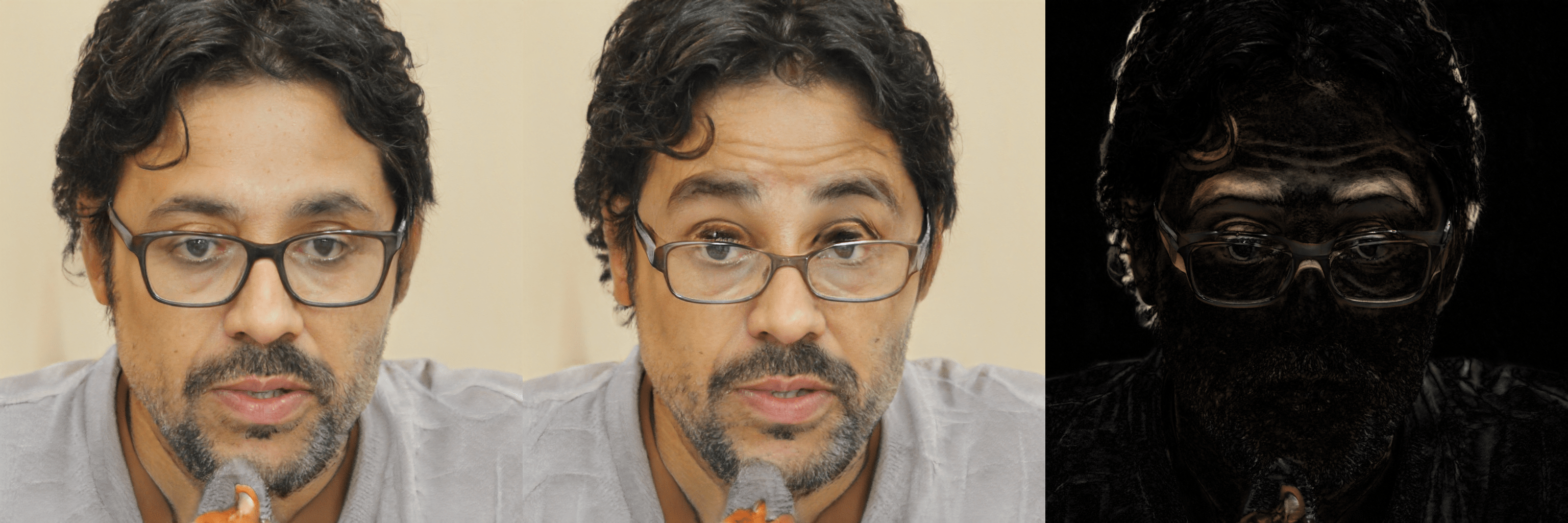}
\caption{Example of Image Input to VLM.}
\label{fig:example-VLM}
\end{figure}

In this section, \tool incorporates a vision-language model (VLM) both as a semantic evaluator and as an automated annotator, as illustrated in steps (d)~Relevance Attribution and (f)~Relabel in \autoref{fig:pipeline}.
The VLM provides textual-aligned supervision for generated images, enabling \tool to automatically determine whether each perturbation expresses a task-relevant concept or a spurious one.

\tool's relevance attribution module systematically categorizes each feature $\phi$ into $\mathcal{F}_{\text{rel}}^\mathcal{T}$ or $\mathcal{F}_{\text{irr}}^\mathcal{T}$ in a data-driven manner.
Specifically, as an example shown in \autoref{fig:example-VLM}, the image input to VLM is composed of three pictures spliced together: (1)~the original image, (2)~the image after a single latent-direction change, and (3)~a difference mask that highlights where the change occurs. The prompt instructs the VLM only to judge changes to the relevant attribute. 
For each channel in the influential feature set, \tool aggregates 
VLM decisions across multiple counterfactual samples and assigns a label by majority voting. 
Channels whose perturbations align with the task semantics are labeled as relevant, while those inducing visually noticeable but task-irrelevant changes are labeled as spurious. 
\changed{For instance, in \autoref{fig:example-VLM}, a perturbation majorly alters the eyebrow region without affecting the presence of glasses; such a change is identified as task-irrelevant.}

Beyond this semantic supervision, \tool also exploits the inherent spatial hierarchy of StyleGAN. Empirically, channels from coarse-to-middle layers—responsible for high-level attributes such as shape, pose, or object identity—tend to correspond to relevant semantics, whereas fine-layer perturbations often lead to superficial changes like lighting or color shifts~\cite{stylegan1-Karras2021}.

Then, for the subset of relevant features, the VLM is employed again to relabel the decision-boundary images produced in the oracle misclassification loop, since the original labels are no longer preserved. For spurious features, \tool restores the original labels, consistent with our assumption of feature invariance. 

Finally, \tool incorporated both relevant and spurious features into a repaired dataset for fine-tuning of SUT to improve both generalizability and robustness. The fine-tuning follows a lightweight strategy using a small learning rate and limited epochs, allowing the model to adapt to corrected feature-label associations without overfitting to the generated data.

\section{Empirical Study}\label{sec:study}

\subsection{Research Questions}
\noindent
\textbf{RQ\textsubscript{1} (configuration):} 
What configuration of \tool yields the most effective and efficient feature-aware testing?

\textbf{RQ\textsubscript{1.1}:} Which sensitivity screening strategy best identifies impactful latent channels? 

\textbf{RQ\textsubscript{1.2}:} What is the optimal number of channels to balance coverage and efficiency?

\textbf{RQ\textsubscript{1.3}:} \changed{How do different VLMs affect the accuracy and cost-efficiency of feature attribution and labeling?}

\noindent
\textbf{RQ\textsubscript{2} (behaviour exploration):} How effective is \tool at finding misbehaviors?   

\noindent
\textbf{RQ\textsubscript{3} (spurious feature detection):} How effective is \tool at discovering failures due to spurious features?

\noindent
\textbf{RQ\textsubscript{4} (model repairing):} Does the generated test cases contribute to improving the SUT?

RQ\textsubscript{1} investigates how to best configure \tool for effective and efficient feature-aware testing. We examine which sensitivity screening strategy most effectively identifies impactful latent channels for targeted perturbation, explore how many channels should be perturbed to balance coverage and efficiency, and analyze how different VLMs influence the accuracy and cost-efficiency of feature attribution and labeling. 
RQ\textsubscript{2} evaluates \tool's capability to uncover behavioral misbehaviors in the system under test (SUT). 
RQ\textsubscript{3} focuses on assessing whether \tool can reveal failures caused by spurious or non-robust features.
Finally, RQ\textsubscript{4} investigates whether the generated test cases can be leveraged to improve the SUT through retraining.

\subsection{Metrics used for Analysis}

For RQ\textsubscript{1.1}, we compare different configurations of \tool in terms of their runtime (in seconds) and the extent to which they alter the SUT's behavior.

\changed{For RQ\textsubscript{1.2}, we measure the number of features in each layer that can affect the behavior of the SUT.}

\changed{For RQ\textsubscript{1.3}, we measure the agreement rate between the VLM and multiple human annotators. 
Besides, to mitigate inconsistency among different annotators, we evaluate consensus agreement against a majority-vote human label, where aggregated labels from multiple annotators represent the consensus ground truth.
In addition to correctness, we also assess the monetary cost per evaluation and average response time to measure their computational and economic efficiency.}

For RQ\textsubscript{2}, we evaluate the runtime of the compared methods in seconds. 
To assess whether structural information is preserved in the generation, we compute the MS-SSIM~\cite{1292216} score, which extends SSIM by computing similarity across multiple scales via iterative downsampling. This multi-scale structure is more in line with human perception than regular SSIM.
Since MS-SSIM is less sensitive to color variation, we complement it with the normalized $\widetilde{d}_2$-Image distance defined as $\widetilde{d}_2(x_i, x_j) = L_2(x_i - x_j)/L_2(\mathbf{1}^x)$, where $\mathbf{1}^x$ is an all-ones matrix with the same shape as $x$.
This metric aims to capture the overall magnitude of change, including color shifts.
Finally, we evaluate the proximity to the decision boundary to assess the effectiveness of the generated test cases in challenging the model's predictions.

For RQ\textsubscript{3},
we propose a simple metric to quantify how much a model relies on task-relevant features, i.e., the proportion of influential channels that are task-relevant: 
\[
R_{Relevance} =\frac{ |\text{Task-Relevant Influential Features}|} { |\text{All Influential Features}|}.
\]
This metric captures the alignment between a model's decision-making process and the task-defined semantics. A higher value indicates that the model's behavior is driven more by meaningful features, whereas a lower value suggests susceptibility to spurious correlations.

\subsection{Objects of Study} 

\subsubsection{Datasets and Models} 
To ensure the generality and robustness of our evaluation, we selected three diverse feature-rich datasets, each paired with a corresponding SUT.

\head{Face Attributes (CelebA~\cite{celeba})} 
This dataset is labeled with 40 facial attributes, with each attribute as an independent binary classification task, although our subsequent analyses primarily focus on the eyeglasses and gender attributes. We evaluate two different SUTs: 
(1)~\textit{ResNet50~\cite{resnet}}, consisting of a pretrained backbone followed by a custom fully connected classification layer. The entire model is fully fine-tuned on CelebA to predict all 40 attributes. It achieves 99.70\% test accuracy on the eyeglasses attribute and 98.17\% on gender. 
(2)~\textit{SWAG ViT~\cite{swag-Mehta-2022}} a Vision Transformer with a frozen pretrained backbone and a custom two-layer classification head, trained in a weakly supervised manner on CelebA attributes.
This model is chosen for its reported robustness against spurious features. It achieves a test accuracy of $99.17\%$ on the eyeglasses attribute and $98.91\%$ on the gender attribute.

\head{Dog breed Classification (Dogs~\cite{dogs_kaggle})}
We use a publicly available dataset from Kaggle~\cite{dogs_kaggle}, which includes labeled images of 78 dog breeds. This dataset is used to evaluate multi-class classification performance. 
Only dog images were retained for our multi-class classification task. The model under test is a pretrained ReXNet-150~\cite{rexnet}, which achieved a test accuracy of $87.9\%$. 

\head{Car Detection (COCO~\cite{coco})} 
We evaluate object detection performance on the car class. The supervised model under test is YOLOv8, implemented via the official Ultralytics framework~\cite{yolo}. The model is pretrained on the COCO dataset, and we directly apply this pretrained model to detect cars in our images without any fine‑tuning. 
On the LSUN Cars dataset~\cite{lsun}, we report an accuracy of $97.63\%$ when including trucks, since some vehicles are ambiguous and may be classified as either car or truck; accuracy considering only cars is $78.83\%$.

\subsubsection{Generator and VLM} 

Each dataset is paired with a pretrained StyleGAN generator and segmentation model. 
We use the official StyleGAN2 models pretrained on FFHQ~\cite{stylegan2-karras2020}, LSUN Dog~\cite{lsun}, and LSUN Car~\cite{lsun}, and further fine-tune each generator on the corresponding downstream dataset using a few thousand task-specific images ($\sim$5k steps) and adaptive discriminator augmentation (ADA). 
This setup follows common practice in latent-space test generation methods~\cite{2025-Maryam-ICST, mozumder2025rbt4dnnrequirementsbasedtestingneural}, where pretrained generators are preferred because their latent spaces capture a wider range of features learned from large and diverse datasets. Directly training a generator from scratch on the same dataset as SUT typically limits the richness of the learned feature space.

\subsubsection{Configurations}

Regarding configuration of sensitive channel screening (RQ\textsubscript{1.1}), we evaluate three methods in \tool, including two white-box approaches, \textit{Gradient Saliency} (\textit{Grad}) and \textit{SmoothGrad}, and one black-box approach, the \textit{Finite-Difference Approximation} (\textit{FDA}). 
We chose these methods for different application requirements and computational complexity. \textit{Grad} requires a single backward pass through the SUT and generator. \textit{SmoothGrad} increases this cost by performing $N$ backward passes, each on a perturbed version of the input, to generate more robust attributions through averaging. Following prior work~\cite{smoothgrad-smilkov2017}, we set $N=10$. 
In contrast, \textit{FDA} is gradient-free and suitable for black-box settings. Instead, it perturbs each individual channel of style vectors and observes the resulting changes. 
Given that our StyleGAN model contains $9,088$ channels, \textit{FDA} performs $9,088$ SUT forward passes to approximate relevance.
\changed{Besides,  to support relevance attribution and relabel procedures (RQ\textsubscript{1.3}), we evaluate 8 VLMs covering different model families, sizes, and reasoning capabilities. The selected models include Qwen-VL~\cite{Qwen} (8B-Thinking, 30B-A3B-Thinking, and Max), Mistral VLM variants~\cite{mistral} (Small-3.1-24b-instruct, Medium-3.1), and OpenAI models~\cite{gpt} (GPT-o4-Mini, GPT-4.1). }

\changed{Regarding RQ\textsubscript{2}, to ensure a fair comparison with conventional test generation methods, we also define a simplified version of our framework, denoted as $\tool^{*}$.
Unlike the full pipeline, $\tool^*$ omits the relevance attribution step and perturbs all channels identified as sensitive in the initial screening. It relies solely on the misclassification-based oracle and thus functions as a general-purpose test generator, targeting exposing misbehaviours under small, single-feature perturbations.
This configuration is used when semantic relevance cannot be reliably established. In certain domains, relevance attribution requires expert-level domain knowledge and cannot be handled reliably by general-purpose VLMs, such as the medical applications or fine-grained species identification (\dogs). In other settings, the generator may be trained exclusively on a specific domain, so all latent variations describe within-class changes, such as \cars. In such cases, distinguishing relevant from spurious features becomes unnecessary since all the detected features are deemed irrelevant. In such cases, $\tool^{*}$ serves as a domain-agnostic variant that allows fair comparison with conventional latent-space test generators.}
\subsubsection{Comparison Baselines} 

Although no existing method jointly addresses the dual objectives of our approach, we compare \tool to representative baselines that align with each individual testing objective, latent-space test input generation, and spurious feature identification

\head{Mimicry} 
A state-of-the-art latent-space test generation method, that exploits the $\mathcal{W}$-space of StyleGAN. It interpolates between $w$-latent vectors from different class instances and uses population-based optimization to tune the interpolation weights. The resulting synthesized images form test cases positioned near the SUT's decision boundary~\cite{mimicry-weißl2025}. Like many other latent-space test generators, the entire latent vector is manipulated during this process, confounding relevant and irrelevant features. In contrast, \tool operates in the more disentangled $\mathcal{S}$-space and efficiently screens sensitive features to the SUT and only perturbs a single semantic feature at a time.
To support a fair comparison, we primarily compare Mimicry against $\tool^*$, the simplified variant of our framework that does not use relevance attribution and perturbs all channels identified as sensitive.
\changed{Mimicry serves as a strong baseline for evaluating generated test cases, and has been shown to surpass earlier methods such as Sinvad~\cite{kang2020sinvad} and DeepJanus~\cite{riccio2020deepjanus} in boundary testing settings.}

\head{SpRAy (spectral relevance analysis)}
An eigenvalue-based clustering method for spurious feature localization proposed by Lapuschkin et al.~\cite{spray_Lapuschkin_2019, anders2022finding}.
It is based on computing Layer-wise Relevance Propagation (LRP) heatmaps for model predictions and then applying spectral clustering to group similar explanation patterns.
We include SpRAy as a baseline to evaluate effectiveness in spurious feature detection, since it remains one of the few techniques that directly target the automatic detection of spurious features. 
SpRAy \changed{represents the pixel-level explanation family of methods and clustering saliency explanations. As such, SpRAy serves as a representative and methodologically relevant baseline for assessing the effectiveness of spurious feature detection from explanation patterns, complementing our feature-perturbation-based analysis.}

\subsection{Procedure}

In RQ\textsubscript{1}, we apply each sensitive channel screening method using the confidence-based oracle on the facial task for 50 randomly selected seeds. 
This is justified as channel sensitivity mainly depends on StyleGAN's internal structure rather than task semantics, making the results broadly transferable and reused across datasets in RQ\textsubscript{2} and RQ\textsubscript{3}.
We fine-tuned the perturbation $\epsilon=10$ to induce visible semantic changes and avoid artifacts, following existing work~\cite{2025-Maryam-ICST,2023-Riccio-ICSE}.

 Additionally, we analyze the contribution of specific synthesis layer channels to justify the selection of the number of candidate channels in subsequent experiments.
\changed{To compute the human-aligned metrics for RQ\textsubscript{1.3}, we collected human annotations using AWS Mechanical Turk~\cite{sorokin2008utility}. Two tasks were evaluated: feature attribution and boundary-image relabeling. In total, 69 participants completed 1,069 annotation questions. We included several attention-check questions to detect inattentive responses and discarded 22 submissions that failed these checks.} 

In RQ\textsubscript{2}, we select the best-performing configuration from RQ\textsubscript{1} against \mimicry. For this comparison, we use the oracle misclassification, which resembles the boundary-based objectives used in \mimicry~\cite{mimicry-weißl2025} and evaluate with 50 seeds by a random number generator (RNG).
Following existing guidelines in generative AI-based test generation~\cite{2025-Maryam-ICST,stylegan2-karras2020}, \tool applies latent truncation to improve image quality. For the face and dogs tasks, we use a truncation value of $\psi = 0.7$, while for cars detection we use $\psi = 0.5$. 
Beyond runtime, we assess the visual impact of manipulations using $d_2$-distance (lower is better) and MS-SSIM~\cite{1292216} (higher indicates greater structural similarity).

In RQ\textsubscript{3}, we evaluate spurious feature detection using 88 RNG seeds for both the ResNet and SWAG models on the eyeglasses classification task \changed{and 50 RNG seeds for YOLO on the car detection task}. We use a  40\% confidence drop threshold to define influential features in the confidence-based oracle, $\tau = 0.4|y[t]|$, to capture impactful shifts in prediction without overreacting to minor fluctuations.
We re-implemented the SpRAy method based on the paper description~\cite{spray_Lapuschkin_2019}, as no official code was provided. The procedure includes the following steps: First, we generate LRP relevance maps for positive samples in the test set. Then, we pre-process the maps by resizing and flattening them into uniform vectors and perform eigengap analysis on the k-nearest neighbor affinity graph to determine the optimal number of clusters. Finally, we apply spectral clustering~\cite{spectral-clustering} and visualize clusters via t-SNE~\cite{t-sne}.
This process serves as a baseline to assess the capability of \tool on identifying spurious correlations.

In RQ\textsubscript{4}, we fine-tune the pretrained CelebA ResNet classifier using a small repaired dataset generated by \tool. More specifically, the repairing dataset mixes original CelebA training samples (80\%) with generated counterfactuals (20\%). 
Only the final classification layer is updated, while the backbone remains frozen. The model is optimized with a small learning rate ($\text{lr}=2\times 10^{-5}$), early stopping for at most 20 epochs, and a loss targeting only the finetuned attribute (eyeglasses), ensuring that the model adapts specifically to corrected feature–label associations. To ensure that evaluation is conducted on data unseen during fine-tuning, we create a held-out generated test set by reserving 20\% of all VLM-relabeled counterfactuals using a fixed random seed.

\subsection{Results}

\subsubsection{RQ\textsubscript{1} (configuration)}
\autoref{tab:rq1_stats} summarizes the average confidence drop and runtime for each sensitive channel screening method, along with their standard deviations. The two white-box methods, \textit{Grad} and \textit{SmoothGrad}, perform similarly in reducing model confidence, while \textit{FDA} is less effective. 
In terms of runtime, \textit{Grad} is the fastest, with \textit{SmoothGrad} incurring only modest additional cost due to the multiple backward passes. \textit{FDA}, by contrast, is slower due to its gradient-free sensitivity estimation.

To better understand how many sensitive channels are required for effective perturbation, we analyze the distribution of activated channels across StyleGAN's synthesis layers in cases where the SUT exhibits a confidence drop greater than 1. 
This threshold removes low-signal noise and retains only semantically impactful perturbations, as it reflects a substantial logit change.

\begin{table}[t]
\caption{RQ\textsubscript{1} (configuration) results \textmd{(\textbf{best} \& \underline{second best})}.} 
\label{tab:rq1_stats}
\resizebox{0.8\columnwidth}{!}{
\begin{tabular}{llll}
\toprule
& \textit{Grad}        & \textit{SmoothGrad}  & \textit{FDA}  \\ \midrule
Confidence Drop $\uparrow$ & $\underline{0.506\pm1.230}$ & $\mathbf{0.581\pm1.287}$  & $0.062\pm0.385$   \\
Runtime (sec) $\downarrow$   & $\mathbf{5.486\pm0.294}$ & $\underline{5.838\pm0.290}$  & $139.530\pm1.164$ \\ 
\bottomrule
\end{tabular}}

\end{table} 

\begin{figure}[h]
\centering
\includegraphics[width=0.7\linewidth]{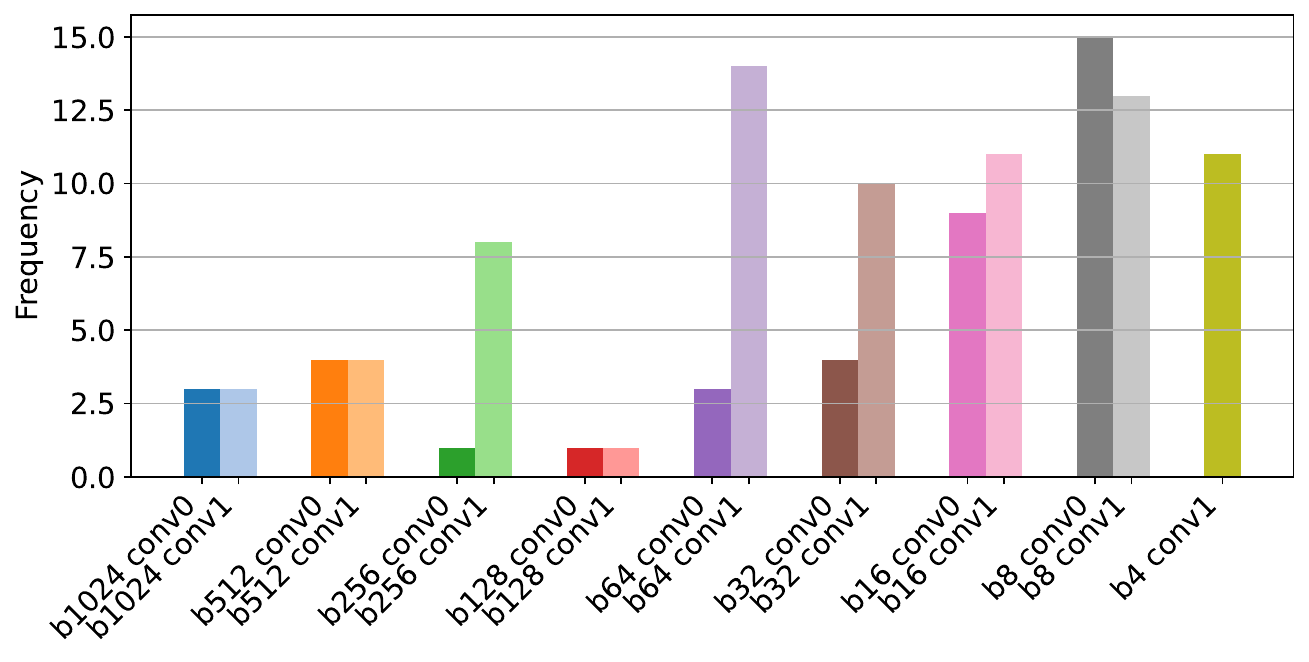}
\caption{RQ\textsubscript{1} (configuration): Number of channels with confidence drop greater than one for each StyleGAN layer in \tool.}
\label{fig:layer_used}
\end{figure}

As shown in \autoref{fig:layer_used}, only a subset of channels, mainly from early layers, are consistently involved in these impactful cases. This analysis suggests that a limited number of candidate channels is sufficient to induce meaningful behavioral shifts while maintaining efficiency. Specifically, we select the top-15 most sensitive channels from the coarse-to-middle layers, and the top-5 from the fine layers (beyond 512 resolution) for subsequent tasks.
While earlier layers are known to control high-level semantics~\cite{stylegan1-Karras2021}, our results confirm that their channels are disproportionately represented among those driving significant output changes.

\begin{figure}[tb]

\includegraphics[width=1.1\linewidth]{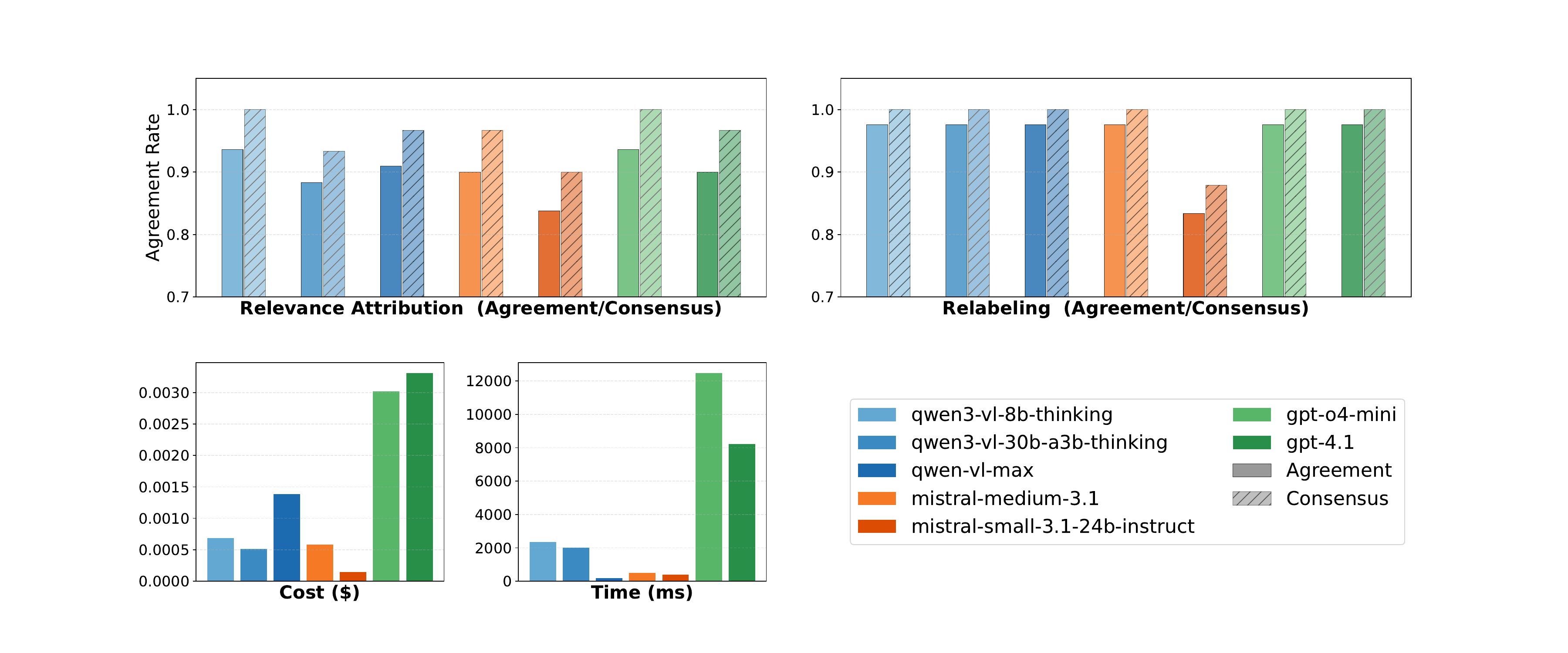}
\caption{RQ\textsubscript{1} (configuration): Performance of VLMs comparison.}
\label{fig:vlms}
\end{figure}
\changed{\autoref{fig:vlms} compares the performance of 7 VLMs across four dimensions.
The top row shows their agreement with human annotators for the two tasks: Relevance Attribution and Relabeling. Each bar pair reports both the raw agreement and the majority-vote consensus agreement.
The Relevance Attribution task is more challenging, where models such as qwen3-vl-8b-thinking and gpt-o4-mini achieve the highest alignment with human judgments.
For the Relabeling task, performance is generally more stable: all models except mistral-small-3.1-24b-instruct attain high agreement levels close to the human upper bound.
The bottom row presents the cost per query and generation time, revealing substantial differences in efficiency: lighter models incur minimal cost and latency, while larger models such as gpt-4.1 exhibit noticeably higher computational overhead.
Considering both semantic alignment and efficiency trade-offs in cost and latency, we select qwen3-vl-8b-thinking as the VLM used in all subsequent experiments.}

\begin{tcolorbox}[boxrule=0pt,sharp corners,boxsep=2pt,left=2pt,right=2pt,top=2.5pt,bottom=2pt]
\textbf{RQ\textsubscript{1} (configuration)}: \textit{
SmoothGrad offers the best overall trade-off between attribution quality and computational cost among sensitivity screening methods. 
Analysis of impactful perturbations reveals that selecting top-15 channels per early/middle layer and top-5 per fine layer balances effectiveness with efficiency and captures most impactful perturbations.
\changed{Across both attribution and relabeling tasks, the evaluated VLMs show high alignment with human judgments, with qwen3-vl-8b-thinking providing the best balance of semantic accuracy and efficiency.}
}
\end{tcolorbox}


\begin{table}[t]
\caption{RQ\textsubscript{2}: Effectiveness and efficiency results.}
\resizebox{0.9\columnwidth}{!}{
\begin{tabular}{clllll}
\toprule
                                &                          & Runtime (sec) $\downarrow$ & $\widetilde{d}_2$-Image $\downarrow$ & MS-SSIM $\uparrow$ & $d_2$-Boundary $\downarrow$ \\ \midrule
\multirow{4}{*}{\makecell{\faces \\ glasses}} & \tool       & $1.288\pm0.354$   & $\mathbf{0.084\pm0.025}$ & $0.730\pm0.054$ & $\mathbf{0.035\pm0.051}$ \\
                                & $\tool^*$                  & $\mathbf{1.256\pm0.384}$   & $0.088\pm0.043$ & $\mathbf{0.731\pm0.053}$ & $0.060\pm0.141$ \\
                                & \mimicry                  & $96.795\pm0.751$  & $0.250\pm0.049$ & $0.420\pm0.073$ & $2.996\pm0.434$ \\
                                & Significance & $8.07\sigma$ \EffBig       & $7.72\sigma$ \EffBig     & $8.03\sigma$ \EffBig     & $8.07\sigma$ \EffBig     \\ \midrule
\multirow{4}{*}{\makecell{\faces \\ gender}}  & \tool         & $\mathbf{1.277\pm0.354}$   & $\mathbf{0.098\pm0.025}$ & $0.668\pm0.090$ & $2.442\pm2.275$ \\
                                & $\tool^*$                   & $3.846\pm2.830$   & $0.115\pm0.083$ & $\mathbf{0.714\pm0.087}$ & $\mathbf{0.566\pm1.109}$ \\
                                & \mimicry                  & $96.342\pm0.676$  & $0.247\pm0.046$ & $0.415\pm0.065$ & $1.867\pm0.624$ \\
                                & Significance & $12.00\sigma$ \EffBig      & $10.26\sigma$ \EffBig    & $11.70\sigma$ \EffBig    & $9.33\sigma$ \EffBig     \\ \midrule
\multirow{3}{*}{\makecell{\textit{COCO} \\ Cars}}           & $\tool^*$                  & $\mathbf{3.247\pm2.208}$   & $\mathbf{0.094\pm0.045}$ & $\mathbf{0.759\pm0.072}$ & $0.644\pm0.038$ \\
                                & \mimicry                & $50.982\pm0.702$  & $0.251\pm0.044$ & $0.332\pm0.083$ & $\mathbf{0.259\pm0.200}$ \\
                                & Significance & $11.50\sigma$ \EffBig      & $11.07\sigma$ \EffBig    & $11.43\sigma$ \EffBig    & $-5.85\sigma $\\ \midrule
\multirow{3}{*}{\dogs}           & $\tool^*$                   & $\mathbf{3.697\pm2.423}$   & $\mathbf{0.073\pm0.043}$ & $\mathbf{0.819\pm0.100}$ & $0.457\pm0.118$ \\ 
                                & \mimicry                  & $12.325\pm0.674$  & $0.281\pm0.065$ & $0.157\pm0.134$ & $\mathbf{0.422\pm0.157}$ \\
                                & Significance & $12.13\sigma$ \EffBig      & $11.82\sigma$ \EffBig    & $11.72\sigma$ \EffBig    & $-0.71\sigma$\\ \bottomrule
\end{tabular}}
\label{tab:combined_rq2}
\end{table}

\subsubsection{RQ\textsubscript{2} (behavior exploration)}
\autoref{tab:combined_rq2} shows that both \tool across all four tasks and three metrics—runtime, $\widetilde{d}_2$-Image distance, and MS-SSIM—$\tool^*$ consistently outperforms \mimicry, which is confirmed by a Mann–Whitney U test~\cite{Wilcoxon1945} ($\sigma > 2$ indicates significance) and Cohen's d effect sizes~\cite{cohen1988statistical} (\EffBig for $d > 1.0$).
For the $d_2$-Boundary metric, results are more mixed. Under $d_2$-Boundary, $\tool^*$ outperforms the baseline on \faces-gender and \faces-glasses, with both \tool variants surpassing \mimicry on \faces-glasses. Here, the better performance of $\tool^*$ against \mimicry is statistically significant. For multi-class tasks, \mimicry performs better in \cars, while for \dogs we could not identify a clearly better approach.

We also provide visual comparison of the generated examples for \tool and \mimicry across all three tasks in Appendix~\ref{app:img-comparison}. These examples illustrate the qualitative differences between the two methods and support the quantitative trends reported above. In particular, \tool generally produces more focused and attribute-specific edits, while \mimicry often introduces broader or incidental changes.

\begin{tcolorbox}[boxrule=0pt,sharp corners,boxsep=2pt,left=2pt,right=2pt,top=2.5pt,bottom=2pt]
\textbf{RQ\textsubscript{2} (behavior exploration)}: \textit{
\tool produces more targeted and visually precise manipulations than \mimicry while preserving structural similarity at a lower runtime, confirming a more controlled and efficient test generation.}
\end{tcolorbox}

\subsubsection{RQ\textsubscript{3} (spurious feature detection)}
\begin{table}[h]
\centering
\caption{RQ\textsubscript{3}: Comparison of spurious feature detection between ResNet and SWAG models. Only frequently used channels (by > 1 input) are selected.}
\label{tab:spurious_detection}
\begin{tabular}{lcc}
\toprule
 & \textbf{ResNet} & \textbf{SWAG} \\
\midrule
Influential inputs (selected/ total) & 415 / 548 & 168 / 524 \\
Channels used (selected / total) & 94 / 236 & 60 / 450 \\
\midrule
Relevant/Spurious inputs & 329/86 & 55/113 \\
Relevant/Spurious channels & 68/36 & 39/56 \\
$R_{Relevance}$ & 0.65 & 0.41 \\
\bottomrule
\end{tabular}
\end{table}

\autoref{tab:spurious_detection} reports the results for spurious feature identification in glasses classification, distinct for a fully fine-tuned ResNet50 and a frozen-backbone SWAG ViT. Although both models achieve high test accuracy on the eyeglasses attribute, their behavior under controlled latent perturbations diverges substantially.

ResNet50 identifies more task-relevant influential inputs (329 vs. 55) and channels (68 vs. 39), whereas SWAG yields more spurious influential inputs (113 vs. 86) and channels (56 vs. 36). The relevance ratio $R_{\text{Relevance}}$ is higher for ResNet (0.65 vs. 0.41), indicating stronger alignment with semantically meaningful features. Despite SWAG's theoretical robustness~\cite{swag-Mehta-2022}, it shows greater reliance on spurious features. These results suggest that SWAG's frozen backbone may limit task adaptation, increasing dependence on incidental cues.

\begin{figure}[t]
\centering
\includegraphics[width=0.8\linewidth]{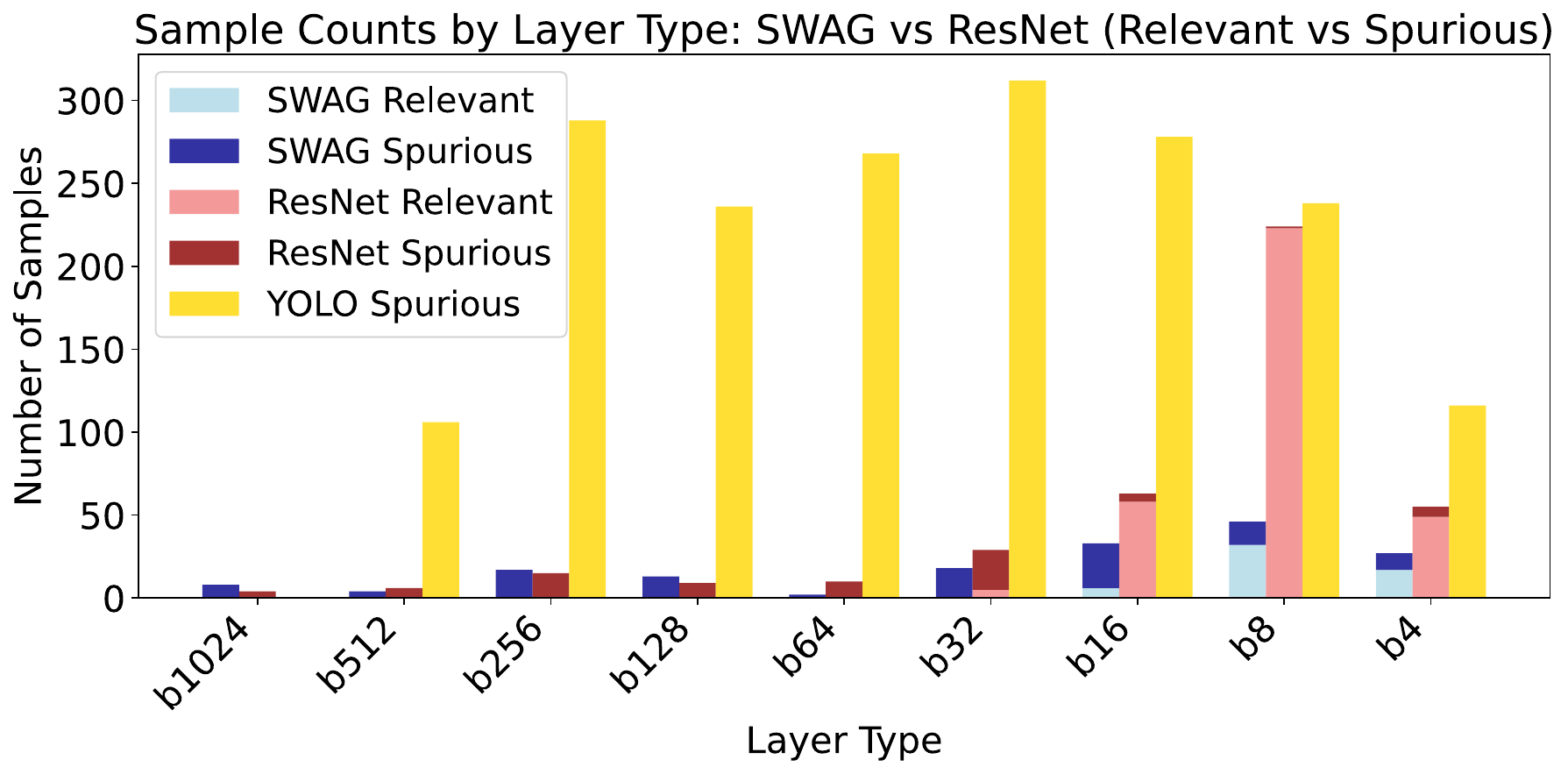}
\caption{RQ\textsubscript{3}: distribution of relevant and spurious attributions across StyleGAN layers for ResNet and SWAG.}
\label{fig:3layer_type_comparison}
\end{figure}

To analyze these results more qualitatively, \autoref{fig:3layer_type_comparison} shows the distribution of relevant and spurious features across StyleGAN layers. Relevant channels concentrate in coarse blocks (b4--b16), which encode semantic structures like shape. In contrast, spurious channels appear more often in mid-to-late layers (e.g., b512 and above), linked to texture and lighting, confirming that many spurious dependencies stem from low-level artifacts.
While ResNet focuses on relevant channels, SWAG exhibits a more dispersed activation pattern with fewer relevant and more spurious attributions, reinforcing the risk of lacking adaptation to relevant features in a weakly supervised manner. Notably, both models' spurious activations concentrate in mid-to-late layers tied to lighting and background, underlining the value of layer-aware attribution for diagnostic insights.
\changed{For YOLO-based car detection, the spurious attributions are spread more uniformly across all layers, unlike the eyeglasses task where they are confined to coarse-level features}

\begin{figure}[h]
\centering
\includegraphics[width=\linewidth]{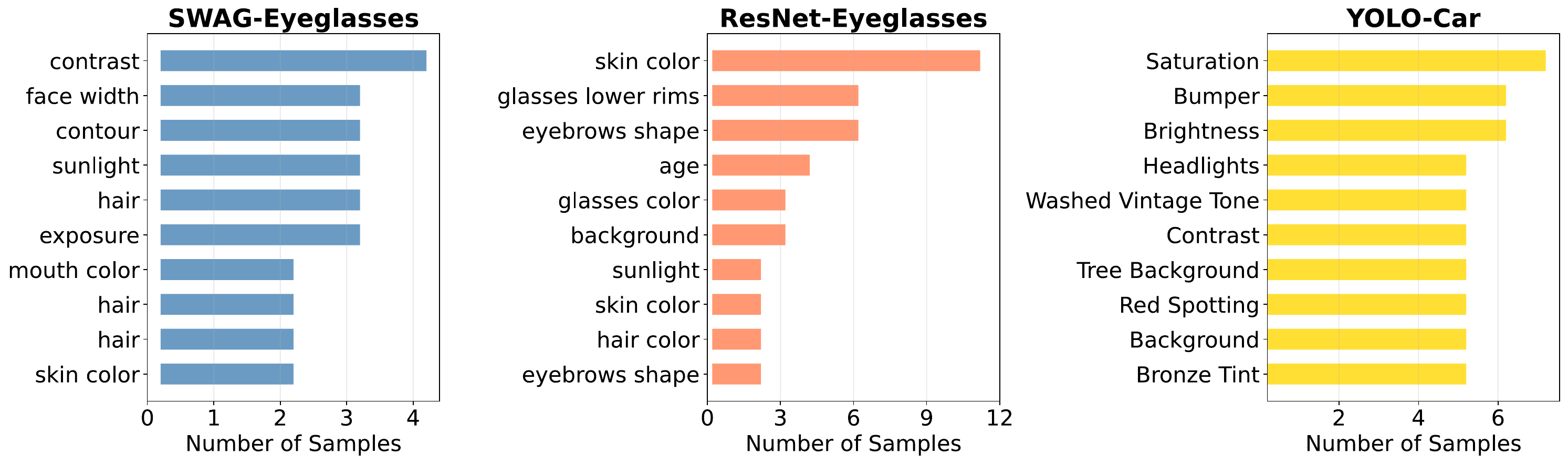}
\caption{RQ\textsubscript{3}: Top-10 spurious features identified for SWAG, ResNet, and YOLO}
\label{fig:top10spurious}
\end{figure}

\autoref{fig:top10spurious} further reveals that SWAG's top spurious directions often reflect global appearance cues, such as face-width, contour, lighting intensity, and exposure. ResNet on the other hand has semantically finer spurious directions, such as glasses frame style, eyebrow height, or clothing. This contrast suggests that ResNet latches onto plausible but misleading features, while SWAG responds more to structural or environmental noise.
\changed{For YOLO car detection, the spurious features span a broader range, including shifts in color saturation, bronze tint, contrast, which indicates that the car detector may rely on a wider mix of context signals than the attribute classifiers.}

\begin{figure}[tb]
\centering
\includegraphics[width=0.7\linewidth]{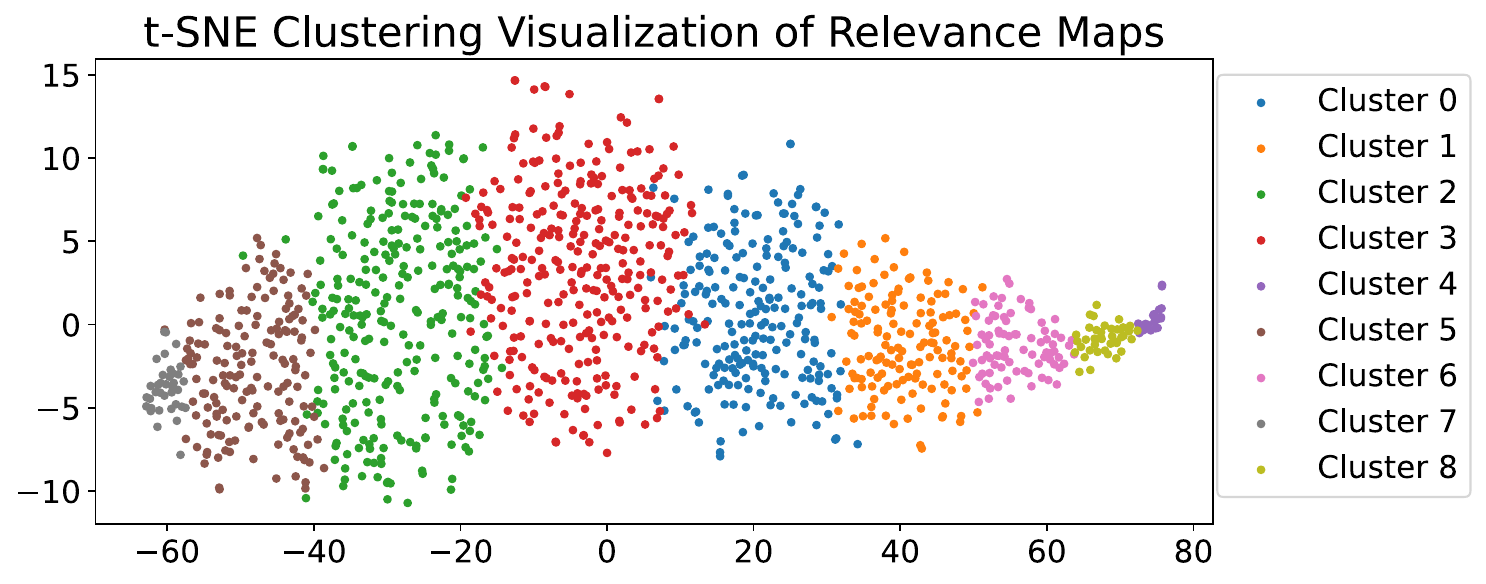}
\includegraphics[width=0.7\linewidth]{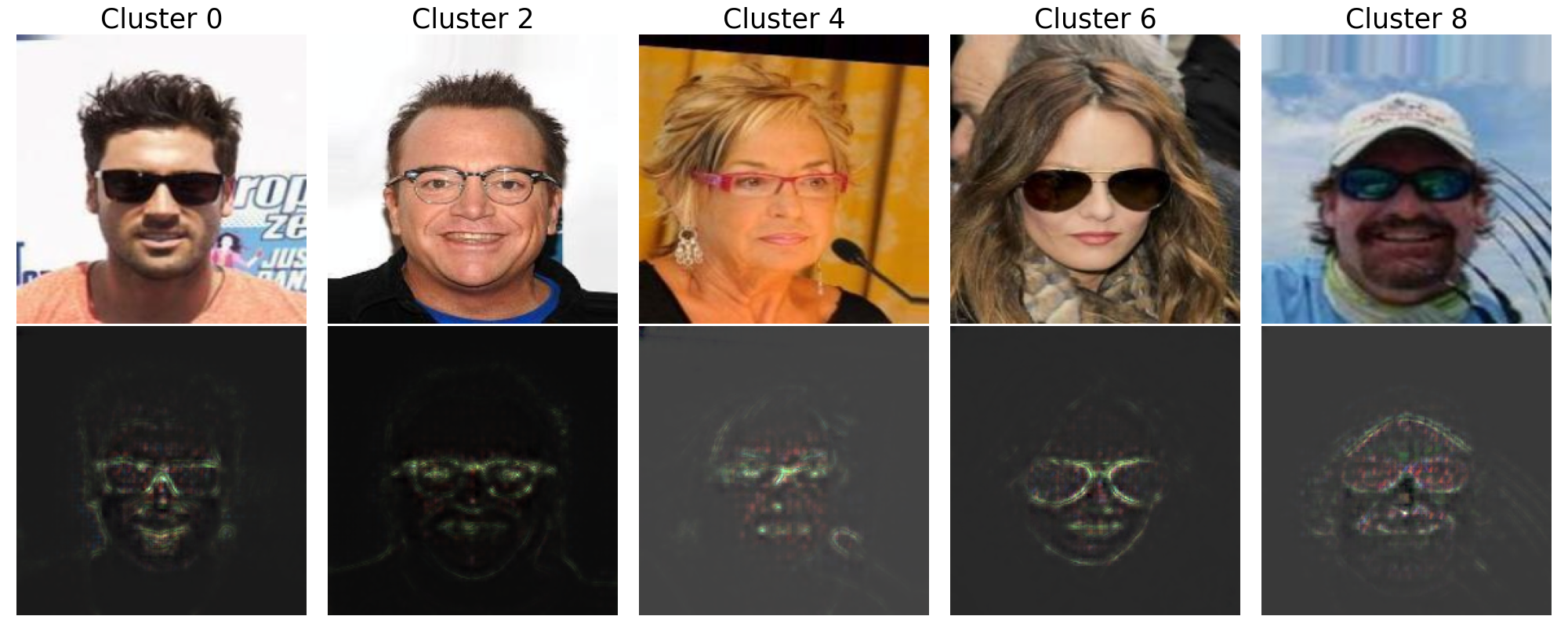}
\caption{RQ\textsubscript{3}: t-SNE visualization of relevance maps by SpRAy. Bottom: samples from five distinct clusters, including original images and corresponding relevance maps.}
\label{fig:spray}
\end{figure}

Finally, \autoref{fig:spray} shows a t-SNE of relevance maps clustered using Spectral Relevance Analysis (SpRAy). While the eigengap analysis suggests 9 clusters, the visualization shows a mostly continuous distribution. Cluster inspection reveals repeated attention around eyeglasses, raising ambiguity over whether these activations reflect causal semantics or spurious correlations. This highlights the limitations of pixel-level relevance methods and the strength of our feature-space probing in uncovering subtle spurious patterns.

\begin{tcolorbox}[boxrule=0pt,sharp corners,boxsep=2pt,left=2pt,right=2pt,top=2.5pt,bottom=2pt]
\textbf{RQ\textsubscript{3} (spurious features detection)}: \textit{
\tool reveals substantial differences in spurious feature reliance between models with similar accuracy, showing that SWAG ViT focuses more on low-level, global cues while ResNet attends to finer but still irrelevant traits. 
 In contrast, SpRAy fails to reliably identify such spurious patterns, especially when irrelevant features co-occur with class semantics.}
\end{tcolorbox}

\subsubsection{RQ\textsubscript{4} (model repair)}

\changed{On the CelebA test set, the fine-tuned model maintains performance comparable to the original model (target accuracy remains ~0.997), indicating that tuning on repaired samples does not degrade performance on standard data.
On the generated boundary-level test set, the improvements are clear: target accuracy increases from 0.935 to 1.000.
These results show that the repaired dataset helps the model handle relevant feature perturbations more reliably while preserving its original behavior.}

\subsection{Threats to Validity}\label{sec:ttv}
\subsubsection{Internal validity}
 The performance of \tool depends on the quality of the pre-trained StyleGAN generator and the accuracy of the semantic segmentation used in feature attribution. Errors in feature relevance estimation or latent disentanglement could lead to incorrect categorization of features as task-relevant or spurious. To mitigate this, we rely on pre-trained models when possible, and apply fine-tuning only when unavoidable, following established guidelines~\cite{stylegan2-karras2020}. Additionally, the threshold selection for sensitivity and logit change might affect the precision of behavioral attribution.

\subsubsection{External validity}
Our evaluation is limited to specific datasets (\faces, \cars, \dogs) and tasks (classification and detection). While we fine-tuned StyleGAN models for better alignment, results may not generalize to domains with less disentangled latent spaces or less structured visual features. Moreover, results on ResNet and SWAG may not extend to all model architectures or training regimes.

\subsubsection{Reproducibility}\label{sec:da}
All software artifacts and the complete results set are available in our replication package~\cite{replication-package}.

\section{Discussion}\label{sec:discussion}

\subsection{From Class- to Feature-Level Testing}
Most prior DL testing methods lack semantic control, making test failures hard to interpret~\cite{riccio2020model,kang2020sinvad}. \mimicry introduced class-conditional testing, improving usability but still limited by the coarser $\mathcal{W}$-space. In contrast, \tool operates in $\mathcal{S}$-space and targets specific semantic features based on model behavior, enabling fine-grained objectives.
As shown in RQ\textsubscript{2}, \tool achieves better boundary discovery on attribute-driven tasks like ``glasses detection'' with higher MS-SSIM and lower $d_2$ distances than \mimicry, indicating subtler, more targeted manipulations. On the dog dataset, both tools perform similarly, though \tool preserves structure better. In the car detection task, \tool is less effective, likely due to the robustness of the object detector requiring coarse-level changes.

A benefit of \tool is its compatibility with unconditional generators as it directly perturbs feature dimensions without needing origin-target class seeds, unlike \mimicry. This quality aspect of \tool is also related to its limitation: selecting only task-relevant channels improves boundary detection for localized attributes (e.g., glasses), but may affect effectiveness on broader concepts like gender, which require more global semantic variation.

\subsection{Spurious Detection vs. Classifier Quality}
The utility of spurious feature detection is closely tied to classifier quality. For low-performing models, such as in fine-grained multi-class dog breed classification, misclassifications frequently stem from underfitting or class confusion rather than shortcut reliance, making spurious attribution less meaningful. 
For these models, test generation is better viewed as a conventional generalization tool. Our framework supports this mode naturally by bypassing feature relevance identification with the more broadly used misclassification oracle. 

Conversely, for high-performing classifiers, accuracy is not sufficient to assess reliability. 
As shown in RQ\textsubscript{3}, both ResNet and SWAG achieve strong accuracy on the glasses task, yet our analysis reveals they rely on different features, including quite distinct spurious patterns.
These findings reinforce the need for feature-aware testing that distinguishes causal semantics from confounding correlations, offering diagnostic insights that go beyond accuracy metrics. To this aim, the proposed metric relevance-based ratio provides a complementary perspective to conventional accuracy.

\subsection{Explainability in Test Generation} 
Incorporating feature-aware perturbations provides a structured form of explainability. Rather than blindly searching for misclassification like \mimicry, \tool distinguishes between task-relevant and irrelevant features, enabling a more informative testing process and more precise judgments about model behavior: failures triggered by irrelevant features reveal spurious reliance (i.e., lack of robustness), while those triggered by relevant ones indicate a lack of generalization.

This fine-grained analysis is particularly important for high-accuracy models, where most inputs are correctly classified. We complement it with additional insights on \textit{why} they are correct. Our framework supports this by converting test generation into a targeted probe of the model's causal reasoning. This not only improves test quality but also enables practical feedback for model assessment, debugging, and retraining.

\subsection{Latent-Space Interpretability} 
Compared to conventional saliency-based methods, our framework introduces a more structured form of explainability grounded in latent space semantics. As shown in RQ\textsubscript{3}, methods like SpRAy produce pixel-level heatmaps that highlight the part where the model focuses, but offer limited insight into what types of features affect the prediction. In contrast, \tool explains model behavior in terms of interpretable latent factors, offering concept-level attribution rather than low-level localization.
\tool bridges the gap between interpretability and explainability: it does not merely visualize what the model reacts to,  but enables the structured reasoning why it reacts that way.

\section{Related Work}\label{sec:related}

\subsection{Test Generation for DNNs}
There are several strategies for generating test inputs for dc learning systems. In this section, we focus on three dominant approaches and contrast them with the capabilities of \tool.

\head{Raw Input Manipulation} 
These methods work directly on the input space, often by perturbing pixel values to induce misbehavior~\cite{szegedy2014intriguingpropertiesneuralnetworks, pei2017deepxplore, guo2018dlfuzz, liu2022deepboundary, croce2020minimally}. Common in adversarial and robustness testing, these techniques apply noise, occlusions, or transformations to existing inputs. While effective, they often produce unrealistic artifacts and are limited in exploring novel input variations~\cite{2023-Riccio-ICSE}. \changed{Since the semantic attributes of the image usually remain unchanged, these methods offer limited insight into how the SUT reacts when meaningful features vary.}

\head{Model-based Techniques} 
They generate test inputs by manipulating high-level, structured representations of the input domain, such as parameterized models encoding domain constraints~\cite{Gambi:2019:ATS:3293882.3330566,riccio2020model,isa}. These methods enable semantically meaningful input synthesis by altering the model and converting it back to raw inputs. However, they rely on manually crafted domain models, which can limit their generalizability~\cite{2023-Riccio-ICSE}.

\head{Generative AI-based Techniques} 
They have recently gained attention for their ability to produce structurally novel test cases without domain models. While Variational Autoencoders (VAEs) have been used~\cite{kang2020sinvad, 2025-Maryam-ICST}, they struggle with complex, context-rich image data~\cite{mimicry-weißl2025}. GANs offer improved generative power~\cite{mimicry-weißl2025, 2025-Maryam-ICST, dola2024cit4dnn}, and recent work is beginning to explore diffusion models as well~\cite{mozumder2025rbt4dnnrequirementsbasedtestingneural, 2025-Maryam-ICST}. \changed{Yet, they usually explore the latent space without distinguishing semantic features. The search in latent space often follows directions that mix attributes or rely on interpolation paths that drift toward invalid regions of the latent manifold, producing artifacts.}

\tool uses a generative AI-based technique, leveraging StyleGAN's architecture for fine-grained control over image features—a capability not yet explored in prior work. It integrates spurious feature detection into the generation process to increase the effectiveness of test cases.

\subsection{Spurious Correlation Detection}
Feature invariance is a foundational principle in robust and fair machine learning, which holds that a model's output should remain unchanged under variations in input features that are semantically irrelevant to the prediction task. This concept has been extensively studied in the context of invariant representation learning and fairness, where models are encouraged to ignore spurious attributes such as background, color, or sensitive demographic variables.

A prominent line of work is \textit{Invariant Risk Minimization}~\cite{invariant-arjovsky2019}, which aims to learn feature representations that support invariant optimal predictors across different environments. Similarly, methods in fairness-aware learning propose counterfactual tests to ensure that models do not change their predictions when protected attributes are altered~\cite{Counterfactual-Kusner2017}. In the context of vision models, feature invariance has been used to detect model biases and to promote robustness against spurious correlations~\cite{Invariance-Ahuja2021}.

Our work builds on this intuition by introducing a test oracle that explicitly evaluates whether a model exhibits invariant behavior with respect to semantically irrelevant features. Unlike prior work that enforces feature invariance during training, our approach uses feature-level perturbations and behavioral analysis to uncover invariance violations in post-hoc testing and diagnosis.

\section{Conclusions}\label{sec:conclusions and Future Work} 

We presented \tool, a feature-aware test generation framework that bridges the gap between test generation and spurious feature diagnosis. By operating in a semantically disentangled latent space, \tool enables fine-grained perturbations aligned with task semantics, and distinguishes between relevant and spurious feature influences. Our experiments show that \tool uncovers behaviorally significant vulnerabilities in high-accuracy models, providing insight into how different models rely on distinct types of features.

As part of our future work, we plan to extend the framework beyond StyleGAN to other generative architectures, investigating how to effectively disentangle—or exploit existing disentangled—latent spaces for feature-aware testing. Another direction is extending the approach to modalities outside vision, such as natural language or audio, where latent structure and semantic drift present related challenges. These directions could broaden the applicability of \tool to a wider range of DL testing.

\balance
\bibliographystyle{ACM-Reference-Format}
\bibliography{paper}

@article{Shortcut-Geirhosetal20,
	title        = {Shortcut learning in deep neural networks},
	author       = {Geirhos, R. and Jacobsen, J.-H. and Michaelis, C. and Zemel, R. and Brendel, W. and Bethge, M. and Wichmann, F. A.},
	year         = 2020,
	journal      = {Nature Machine Intelligence},
	volume       = 2,
	number       = 11,
	pages        = {665--673},
	slug         = {geirhosetal20}
}

@inproceedings{ERM-Vapnik1991,
	title        = {Principles of risk minimization for learning theory},
	author       = {Vapnik, V.},
	year         = 1991,
	booktitle    = {Proceedings of the 5th International Conference on Neural Information Processing Systems},
	location     = {Denver, Colorado},
	publisher    = {Morgan Kaufmann Publishers Inc.},
	address      = {San Francisco, CA, USA},
	series       = {NIPS'91},
	pages        = {831–838},
	isbn         = 1558602224,
	numpages     = 8
}

@INPROCEEDINGS{5206848,
  author={Deng, Jia and Dong, Wei and Socher, Richard and Li, Li-Jia and Kai Li and Li Fei-Fei},
  booktitle={2009 IEEE Conference on Computer Vision and Pattern Recognition}, 
  title={ImageNet: A large-scale hierarchical image database}, 
  year={2009},
  volume={},
  number={},
  pages={248-255},
  doi={10.1109/CVPR.2009.5206848}}

@inproceedings{spurious-training-han2024,
	title        = {Improving group robustness on spurious correlation requires preciser group inference},
	author       = {Han, Yujin and Zou, Difan},
	year         = 2024,
	booktitle    = {Proceedings of the 41st International Conference on Machine Learning},
	location     = {Vienna, Austria},
	publisher    = {JMLR.org},
	series       = {ICML'24},
	articleno    = 696,
	numpages     = 25
}

@inproceedings{gradcam-Selvaraju-2017,
	title        = {Grad-CAM: Visual Explanations from Deep Networks via Gradient-Based Localization},
	author       = {Selvaraju, Ramprasaath R. and Cogswell, Michael and Das, Abhishek and Vedantam, Ramakrishna and Parikh, Devi and Batra, Dhruv},
	year         = 2017,
	booktitle    = {2017 IEEE International Conference on Computer Vision (ICCV)},
	pages        = {618--626},
	doi          = {10.1109/ICCV.2017.74}
}

@misc{spray_Lapuschkin_2019,
	title        = {Unmasking Clever Hans predictors and assessing what machines really learn},
	author       = {Lapuschkin, S. and Wäldchen, S. and Binder, A. and Montavon, G. and Samek, W. and Müller, K.-R.},
	year         = 2019,
	doi          = {10.1038/s41467-019-08987-4},
	url          = {https://publica.fraunhofer.de/handle/publica/258343},
	language     = {en}
}

@article{anders2022finding,
  title={Finding and removing clever hans: Using explanation methods to debug and improve deep models},
  author={Anders, Christopher J and Weber, Leander and Neumann, David and Samek, Wojciech and M{\"u}ller, Klaus-Robert and Lapuschkin, Sebastian},
  journal={Information Fusion},
  volume={77},
  pages={261--295},
  year={2022},
  publisher={Elsevier}
}

@article{lrp-Bach-2015,
	title        = {On pixel-wise explanations for non-linear classifier decisions by layer-wise relevance propagation},
	author       = {Bach, Sebastian and Binder, Alexander and Montavon, Gr{\'e}goire and Klauschen, Frederick and M{\"u}ller, Klaus-Robert and Samek, Wojciech},
	year         = 2015,
	journal      = {PloS one},
	publisher    = {Public Library of Science San Francisco, CA USA},
	volume       = 10,
	number       = 7,
	pages        = {e0130140}
}

@unknown{swag-Mehta-2022,
	title        = {You Only Need a Good Embeddings Extractor to Fix Spurious Correlations},
	author       = {Mehta, Raghav and Albiero, Vítor and Chen, Li and Evtimov, Ivan and Glaser, Tamar and Li, Zhiheng and Hassner, Tal},
	year         = 2022,
	month        = 12,
	doi          = {10.48550/arXiv.2212.06254}
}

@inproceedings{sspace-Wu2021,
	title        = {StyleSpace Analysis: Disentangled Controls for StyleGAN Image Generation},
	author       = {Wu, Zongze and Lischinski, Dani and Shechtman, Eli},
	year         = 2021,
	booktitle    = {2021 IEEE/CVF Conference on Computer Vision and Pattern Recognition (CVPR)},
	pages        = {12858--12867},
	doi          = {10.1109/CVPR46437.2021.01267}
}

@inproceedings{2023-Riccio-ICSE,
	title        = {When and Why Test Generators for Deep Learning Produce Invalid Inputs: an Empirical Study},
	author       = {Vincenzo Riccio and Paolo Tonella},
	year         = 2023,
	booktitle    = {Proceedings of 45th International Conference on Software Engineering},
	publisher    = {ACM},
	series       = {ICSE '23},
	pages        = {12 pages},
	abbr         = {ICSE}
}

@inproceedings{2025-Maryam-ICST,
	title        = {{Benchmarking Generative AI Models for Deep Learning Test Input Generation}},
	author       = {Maryam and Matteo Biagiola and Andrea Stocco and Vincenzo Riccio},
	year         = 2025,
	booktitle    = {Proceedings of 18th IEEE International Conference on Software Testing, Verification and Validation},
	publisher    = {IEEE},
	series       = {ICST '25},
	pages        = {12 pages}
}

@article{stylegan1-Karras2021,
	title        = {{ A Style-Based Generator Architecture for Generative Adversarial Networks }},
	author       = {Karras, Tero and Laine, Samuli and Aila, Timo},
	year         = 2021,
	month        = dec,
	journal      = {IEEE Transactions on Pattern Analysis \& Machine Intelligence},
	publisher    = {IEEE Computer Society},
	address      = {Los Alamitos, CA, USA},
	volume       = 43,
	number       = 12,
	pages        = {4217--4228},
	doi          = {10.1109/TPAMI.2020.2970919},
	issn         = {1939-3539},
	url          = {https://doi.ieeecomputersociety.org/10.1109/TPAMI.2020.2970919}
}

@article{Wilcoxon1945,
	title        = {Individual Comparisons by Ranking Methods},
	author       = {Frank Wilcoxon},
	year         = 1945,
	month        = dec,
	journal      = {Biometrics Bulletin},
	publisher    = {{JSTOR}},
	volume       = 1,
	number       = 6,
	pages        = 80,
	url          = {https://doi.org/10.2307/3001968}
}

@book{cohen1988statistical,
	title        = {Statistical power analysis for the behavioral sciences},
	author       = {Cohen, Jacob},
	year         = 1988,
	publisher    = {L. Erlbaum Associates},
	isbn         = {978-1-134-74270-7}
}

@article{t-sne,
  author  = {Laurens van der Maaten and Geoffrey Hinton},
  title   = {Visualizing Data using t-SNE},
  journal = {Journal of Machine Learning Research},
  year    = {2008},
  volume  = {9},
  number  = {86},
  pages   = {2579--2605},
  url     = {http://jmlr.org/papers/v9/vandermaaten08a.html}
}

@inproceedings{spectral-clustering,
author = {Ng, Andrew Y. and Jordan, Michael I. and Weiss, Yair},
title = {On spectral clustering: analysis and an algorithm},
year = {2001},
publisher = {MIT Press},
address = {Cambridge, MA, USA},
booktitle = {Proceedings of the 15th International Conference on Neural Information Processing Systems: Natural and Synthetic},
pages = {849–856},
numpages = {8},
location = {Vancouver, British Columbia, Canada},
series = {NIPS'01}
}

@inproceedings{riccio2020model,
	title        = {Model-based exploration of the frontier of behaviours for deep learning system testing},
	author       = {Riccio, Vincenzo and Tonella, Paolo},
	year         = 2020,
	booktitle    = {Proceedings of the 28th ACM Joint Meeting on European Software Engineering Conference and Symposium on the Foundations of Software Engineering},
	pages        = {876--888}
}

@article{isa,
	title        = {Identifying and Explaining Safety-critical Scenarios for Autonomous Vehicles via Key Features},
	author       = {Neelofar, Neelofar and Aleti, Aldeida},
	year         = 2024,
	journal      = {ACM Transactions on Software Engineering and Methodology},
	publisher    = {ACM New York, NY},
	volume       = 33,
	number       = 4,
	pages        = {1--32}
}

@inproceedings{stylegan2-karras2020,
	title        = {{ Analyzing and Improving the Image Quality of StyleGAN }},
	author       = {Karras, Tero and Laine, Samuli and Aittala, Miika and Hellsten, Janne and Lehtinen, Jaakko and Aila, Timo},
	year         = 2020,
	month        = {Jun},
	booktitle    = {2020 IEEE/CVF Conference on Computer Vision and Pattern Recognition (CVPR)},
	publisher    = {IEEE Computer Society},
	address      = {Los Alamitos, CA, USA},
	pages        = {8107--8116},
	doi          = {10.1109/CVPR42600.2020.00813},
	url          = {https://doi.ieeecomputersociety.org/10.1109/CVPR42600.2020.00813}
}

@inproceedings{stylegan3-Karras2021,
	title        = {Alias-Free Generative Adversarial Networks},
	author       = {Tero Karras and Miika Aittala and Samuli Laine and Erik H\"ark\"onen and Janne Hellsten and Jaakko Lehtinen and Timo Aila},
	year         = 2021,
	booktitle    = {Proc. NeurIPS}
}

@article{invariant-arjovsky2019,
	title        = {Invariant risk minimization},
	author       = {Arjovsky, Martin and Bottou, L{\'e}on and Gulrajani, Ishaan and Lopez-Paz, David},
	year         = 2019,
	journal      = {arXiv preprint arXiv:1907.02893}
}

@inproceedings{Invariance-Ahuja2021,
	title        = {Invariance principle meets information bottleneck for out-of-distribution generalization},
	author       = {Ahuja, Kartik and Caballero, Ethan and Zhang, Dinghuai and Gagnon-Audet, Jean-Christophe and Bengio, Yoshua and Mitliagkas, Ioannis and Rish, Irina},
	year         = 2021,
	booktitle    = {Proceedings of the 35th International Conference on Neural Information Processing Systems},
	publisher    = {Curran Associates Inc.},
	address      = {Red Hook, NY, USA},
	series       = {NIPS '21},
	isbn         = 9781713845393,
	articleno    = 263,
	numpages     = 13
}

@inproceedings{Counterfactual-Kusner2017,
	title        = {Counterfactual fairness},
	author       = {Kusner, Matt and Loftus, Joshua and Russell, Chris and Silva, Ricardo},
	year         = 2017,
	booktitle    = {Proceedings of the 31st International Conference on Neural Information Processing Systems},
	location     = {Long Beach, California, USA},
	publisher    = {Curran Associates Inc.},
	address      = {Red Hook, NY, USA},
	series       = {NIPS'17},
	pages        = {4069–4079},
	isbn         = 9781510860964,
	numpages     = 11
}

@article{smoothgrad-smilkov2017,
	title        = {Smoothgrad: removing noise by adding noise},
	author       = {Smilkov, Daniel and Thorat, Nikhil and Kim, Been and Vi{\'e}gas, Fernanda and Wattenberg, Martin},
	year         = 2017,
	journal      = {arXiv preprint arXiv:1706.03825}
}

@misc{mimicry-weißl2025,
	title        = {Targeted Deep Learning System Boundary Testing},
	author       = {Oliver Weißl and Amr Abdellatif and Xingcheng Chen and Giorgi Merabishvili and Vincenzo Riccio and Severin Kacianka and Andrea Stocco},
	year         = 2025,
	url          = {https://arxiv.org/abs/2408.06258},
	eprint       = {2408.06258},
	archiveprefix = {arXiv},
	primaryclass = {cs.SE}
}

@inproceedings{zohdinasab2021deephyperion,
	title        = {Deephyperion: exploring the feature space of deep learning-based systems through illumination search},
	author       = {Zohdinasab, Tahereh and Riccio, Vincenzo and Gambi, Alessio and Tonella, Paolo},
	year         = 2021,
	booktitle    = {Proceedings of the 30th ACM SIGSOFT International Symposium on Software Testing and Analysis},
	publisher    = {Association for Computing Machinery},
	series       = {ISSTA '21},
	pages        = {79--90}
}

@inproceedings{pei2017deepxplore,
	title        = {Deepxplore: Automated whitebox testing of deep learning systems},
	author       = {Pei, Kexin and Cao, Yinzhi and Yang, Junfeng and Jana, Suman},
	year         = 2017,
	month        = {oct},
	journal      = {Commun. ACM},
	booktitle    = {proceedings of the 26th Symposium on Operating Systems Principles},
	location     = {Shanghai, China},
	publisher    = {ACM},
	address      = {New York, NY, USA},
	series       = {SOSP '17},
	volume       = 62,
	number       = 11,
	pages        = {1--18},
	doi          = {10.1145/3132747.3132785},
	isbn         = {978-1-4503-5085-3},
	issn         = {0001-0782},
	url          = {http://doi.acm.org/10.1145/3132747.3132785},
	acmid        = 3132785,
	numpages     = 18,
	issue_date   = {November 2019}
}

@inproceedings{guo2018dlfuzz,
	title        = {Dlfuzz: Differential fuzzing testing of deep learning systems},
	author       = {Guo, Jianmin and Jiang, Yu and Zhao, Yue and Chen, Quan and Sun, Jiaguang},
	year         = 2018,
	booktitle    = {Proceedings of the 2018 26th ACM Joint Meeting on European Software Engineering Conference and Symposium on the Foundations of Software Engineering},
	pages        = {739--743}
}

@inproceedings{kang2020sinvad,
	title        = {Sinvad: Search-based image space navigation for dnn image classifier test input generation},
	author       = {Kang, Sungmin and Feldt, Robert and Yoo, Shin},
	year         = 2020,
	booktitle    = {Proceedings of the IEEE/ACM 42nd International Conference on Software Engineering Workshops},
	pages        = {521--528}
}

@inproceedings{liu2022deepboundary,
	title        = {DeepBoundary: A Coverage Testing Method of Deep Learning Software based on Decision Boundary Representation},
	author       = {Liu, Yue and Feng, Lichao and Wang, Xingya and Zhang, Shiyu},
	year         = 2022,
	booktitle    = {2022 IEEE 22nd International Conference on Software Quality, Reliability, and Security Companion (QRS-C)},
	pages        = {166--172},
	organization = {IEEE}
}

@inproceedings{croce2020minimally,
	title        = {Minimally distorted adversarial examples with a fast adaptive boundary attack},
	author       = {Croce, Francesco and Hein, Matthias},
	year         = 2020,
	booktitle    = {International conference on machine learning},
	pages        = {2196--2205},
	organization = {PMLR}
}

@misc{replication-package,
	title        = {Replication Package.},
	year         = 2025,
	howpublished = {\url{https://github.com/Keith-XC/DETECT/}}
}

@misc{szegedy2014intriguingpropertiesneuralnetworks,
	title        = {Intriguing properties of neural networks},
	author       = {Christian Szegedy and Wojciech Zaremba and Ilya Sutskever and Joan Bruna and Dumitru Erhan and Ian Goodfellow and Rob Fergus},
	year         = 2014,
	url          = {https://arxiv.org/abs/1312.6199},
	eprint       = {1312.6199},
	archiveprefix = {arXiv},
	primaryclass = {cs.CV}
}

@book{fda,
author = {Morton, K. W. and Mayers, D. F.},
title = {Numerical Solution of Partial Differential Equations: An Introduction},
year = {2005},
isbn = {0521607930},
publisher = {Cambridge University Press},
address = {USA}
}

@inproceedings{dola2024cit4dnn,
	title        = {CIT4DNN: Generating Diverse and Rare Inputs for Neural Networks Using Latent Space Combinatorial Testing},
	author       = {Dola, Swaroopa and McDaniel, Rory and Dwyer, Matthew B and Soffa, Mary Lou},
	year         = 2024,
	booktitle    = {Proceedings of the IEEE/ACM 46th International Conference on Software Engineering},
	pages        = {1--13}
}

@inproceedings{1292216,
	title        = {Multiscale structural similarity for image quality assessment},
	author       = {Wang, Z. and Simoncelli, E.P. and Bovik, A.C.},
	year         = 2003,
	booktitle    = {The Thrity-Seventh Asilomar Conference on Signals, Systems \& Computers, 2003},
	volume       = 2,
	pages        = {1398--1402 Vol.2},
	doi          = {10.1109/ACSSC.2003.1292216}
}

@misc{resnet,
	title        = {{Deep Residual Learning for Image Recognition}},
	author       = {Kaiming He and Xiangyu Zhang and Shaoqing Ren and Jian Sun},
	year         = 2015,
	booktitle    = {2016 IEEE Conference on Computer Vision and Pattern Recognition (CVPR)},
	pages        = {770--778},
	doi          = {10.1109/CVPR.2016.90},
	eprint       = {1512.03385},
	archiveprefix = {arXiv},
	primaryclass = {cs.CV}
}

@misc{rexnet,
	title        = {Rethinking Channel Dimensions for Efficient Model Design},
	author       = {Dongyoon Han and Sangdoo Yun and Byeongho Heo and YoungJoon Yoo},
	year         = 2021,
	eprint       = {2007.00992},
	archiveprefix = {arXiv},
	primaryclass = {cs.CV}
}

@misc{dogs_kaggle,
	title        = {Dogs and Cats Classifier Dataset},
	author       = {RoboFlow User ldrago},
	year         = 2022,
	url          = {https://www.kaggle.com/datasets/rajarshi2712/dogs-and-cats-classifier},
	howpublished = {Kaggle}
}

@software{yolo,
	title        = {Ultralytics YOLO},
	author       = {Glenn Jocher and Jing Qiu and Ayush Chaurasia},
	year         = 2023,
	url          = {https://ultralytics.com},
	note         = {If you use this software, please cite it using this metadata. Released under AGPL-3.0. Repository: \url{https://github.com/ultralytics/ultralytics}},
	version      = {8.0.0},
	date         = {2023-01-10}
}

@inproceedings{coco,
	title        = {Microsoft COCO: Common Objects in Context},
	author       = {Lin, Tsung-Yi and Maire, Michael and Belongie, Serge and Hays, James and Perona, Pietro and Ramanan, Deva and Doll{\'a}r, Piotr and Zitnick, C. Lawrence},
	year         = 2014,
	booktitle    = {European Conference on Computer Vision (ECCV)},
	publisher    = {Springer},
	pages        = {740--755},
	doi          = {10.1007/978-3-319-10602-1_48}
}

@inproceedings{celeba,
	title        = {Deep Learning Face Attributes in the Wild},
	author       = {Liu, Ziwei and Luo, Ping and Wang, Xiaogang and Tang, Xiaoou},
	year         = 2015,
	month        = {December},
	booktitle    = {Proceedings of International Conference on Computer Vision (ICCV)}
}

@article{lsun,
	title        = {LSUN: Construction of a Large-scale Image Dataset using Deep Learning with Humans in the Loop},
	author       = {Yu, Fisher and Zhang, Yinda and Song, Shuran and Seff, Ari and Xiao, Jianxiong},
	year         = 2015,
	journal      = {arXiv preprint arXiv:1506.03365}
}

@inproceedings{Gambi:2019:ATS:3293882.3330566,
	title        = {Automatically Testing Self-driving Cars with Search-based Procedural Content Generation},
	author       = {Gambi, Alessio and Mueller, Marc and Fraser, Gordon},
	year         = 2019,
	booktitle    = {Proceedings of the 28th ACM SIGSOFT International Symposium on Software Testing and Analysis},
	location     = {Beijing, China},
	publisher    = {ACM},
	address      = {New York, NY, USA},
	series       = {ISSTA 2019},
	pages        = {318--328},
	doi          = {10.1145/3293882.3330566},
	isbn         = {978-1-4503-6224-5},
	url          = {http://doi.acm.org/10.1145/3293882.3330566},
	acmid        = 3330566,
	numpages     = 11
}

@misc{mozumder2025rbt4dnnrequirementsbasedtestingneural,
      title={RBT4DNN: Requirements-based Testing of Neural Networks}, 
      author={Nusrat Jahan Mozumder and Felipe Toledo and Swaroopa Dola and Matthew B. Dwyer},
      year={2025},
      eprint={2504.02737},
      archivePrefix={arXiv},
      primaryClass={cs.SE},
      url={https://arxiv.org/abs/2504.02737}, 
}

@article{saliency_simonyan2013deep,
	title        = {Deep inside convolutional networks: Visualising image classification models and saliency maps},
	author       = {Simonyan, Karen and Vedaldi, Andrea and Zisserman, Andrew},
	year         = 2013,
	journal      = {arXiv preprint arXiv:1312.6034}
}

@article{Qwen,
  title={Qwen2.5-VL Technical Report},
  author={Bai, Shuai and Chen, Keqin and Liu, Xuejing and Wang, Jialin and Ge, Wenbin and Song, Sibo and Dang, Kai and Wang, Peng and Wang, Shijie and Tang, Jun and others},
  journal={arXiv preprint arXiv:2502.13923},
  year={2025}
}

@misc{mistral,
      title={Mistral 7B}, 
      author={Albert Q. Jiang and Alexandre Sablayrolles and Arthur Mensch and Chris Bamford and Devendra Singh Chaplot and Diego de las Casas and Florian Bressand and Gianna Lengyel and Guillaume Lample and Lucile Saulnier and Lélio Renard Lavaud and Marie-Anne Lachaux and Pierre Stock and Teven Le Scao and Thibaut Lavril and Thomas Wang and Timothée Lacroix and William El Sayed},
      year={2023},
      eprint={2310.06825},
      archivePrefix={arXiv},
      primaryClass={cs.CL},
      url={https://arxiv.org/abs/2310.06825}, 
}

@article{gpt,
  title={Gpt-4 technical report},
  author={Achiam, Josh and Adler, Steven and Agarwal, Sandhini and Ahmad, Lama and Akkaya, Ilge and Aleman, Florencia Leoni and Almeida, Diogo and Altenschmidt, Janko and Altman, Sam and Anadkat, Shyamal and others},
  journal={arXiv preprint arXiv:2303.08774},
  year={2023}
}

@inproceedings{sorokin2008utility,
	title        = {Utility data annotation with amazon mechanical turk},
	author       = {Sorokin, Alexander and Forsyth, David},
	year         = 2008,
	booktitle    = {2008 IEEE computer society conference on computer vision and pattern recognition workshops},
	pages        = {1--8},
	organization = {IEEE}
}

@article{riccio2020testing,
	title        = {Testing machine learning based systems: a systematic mapping},
	author       = {Riccio, Vincenzo and Jahangirova, Gunel and Stocco, Andrea and Humbatova, Nargiz and Weiss, Michael and Tonella, Paolo},
	year         = 2020,
	journal      = {Empirical Software Engineering},
	publisher    = {Springer},
	volume       = 25,
	pages        = {5193--5254}
}

@article{zhang2020machine,
	title        = {Machine learning testing: Survey, landscapes and horizons},
	author       = {Zhang, Jie M and Harman, Mark and Ma, Lei and Liu, Yang},
	year         = 2020,
	journal      = {IEEE Transactions on Software Engineering},
	publisher    = {IEEE},
	volume       = 48,
	number       = 1,
	pages        = {1--36},
	doi          = {10.1109/TSE.2019.2962027}
}

@inproceedings{riccio2020deepjanus,
	title        = {Model-based exploration of the frontier of behaviours for deep learning system testing},
	author       = {Riccio, Vincenzo and Tonella, Paolo},
	year         = 2020,
	booktitle    = {Proceedings of the 28th ACM Joint Meeting on European Software Engineering Conference and Symposium on the Foundations of Software Engineering},
	pages        = {876--888}
}

\clearpage
\appendix
\section{Generated Examples}
\label{app:img-comparison}
\begin{figure}[h]
    \centering

    \begin{subfigure}[b]{0.7\columnwidth}
        \centering
        \includegraphics[width=0.95\columnwidth]{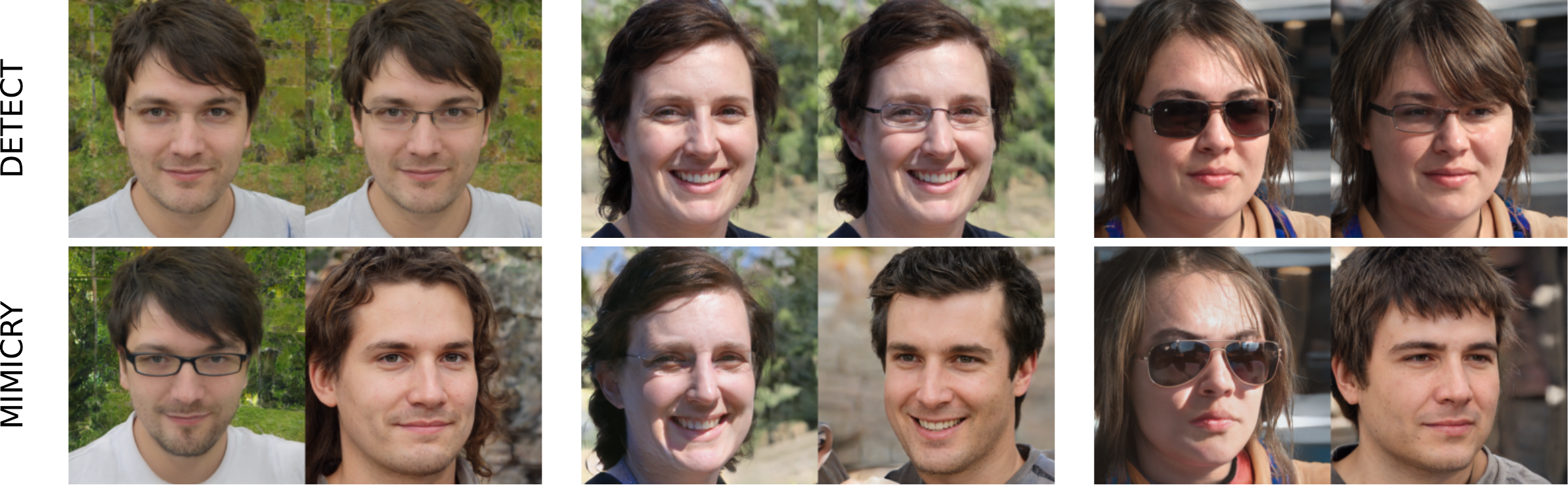}
        \caption{\faces-glasses Task: Change whether Glasses are worn.}
        \label{fig:sub1}
    \end{subfigure}

    \begin{subfigure}[b]{0.7\columnwidth}
        \centering
        \includegraphics[width=0.95\columnwidth]{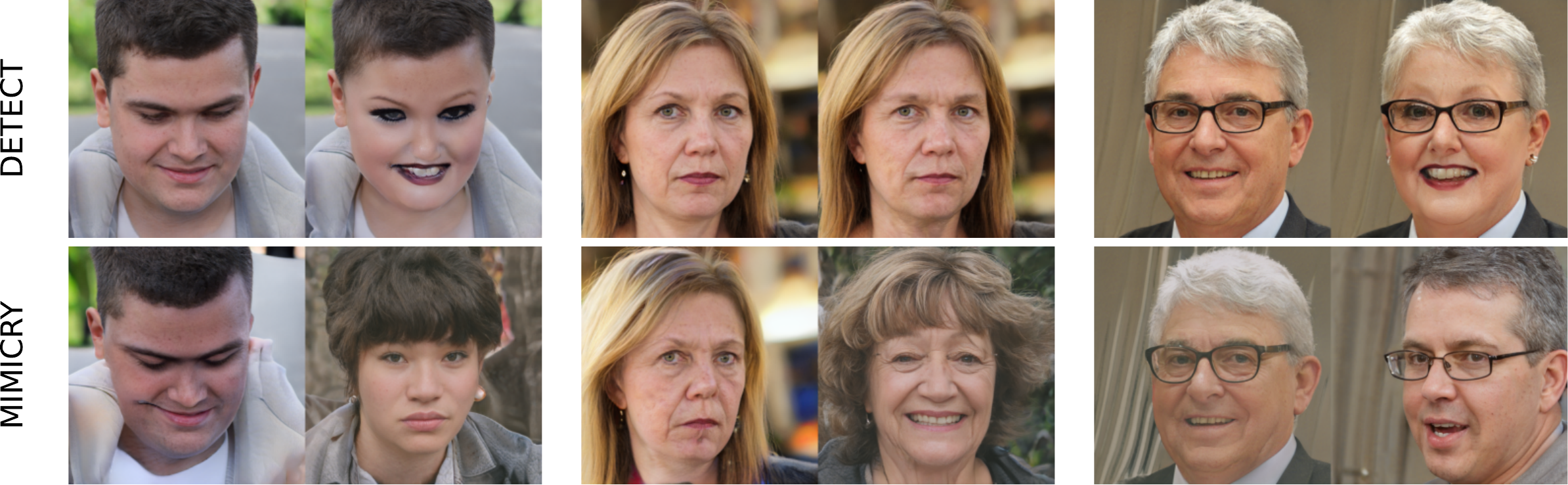}
        \caption{\faces-gender Task: Change Gender of the Person.}
        \label{fig:sub2}
    \end{subfigure}

    \begin{subfigure}[b]{0.7\columnwidth}
        \centering
        \includegraphics[width=0.95\columnwidth]{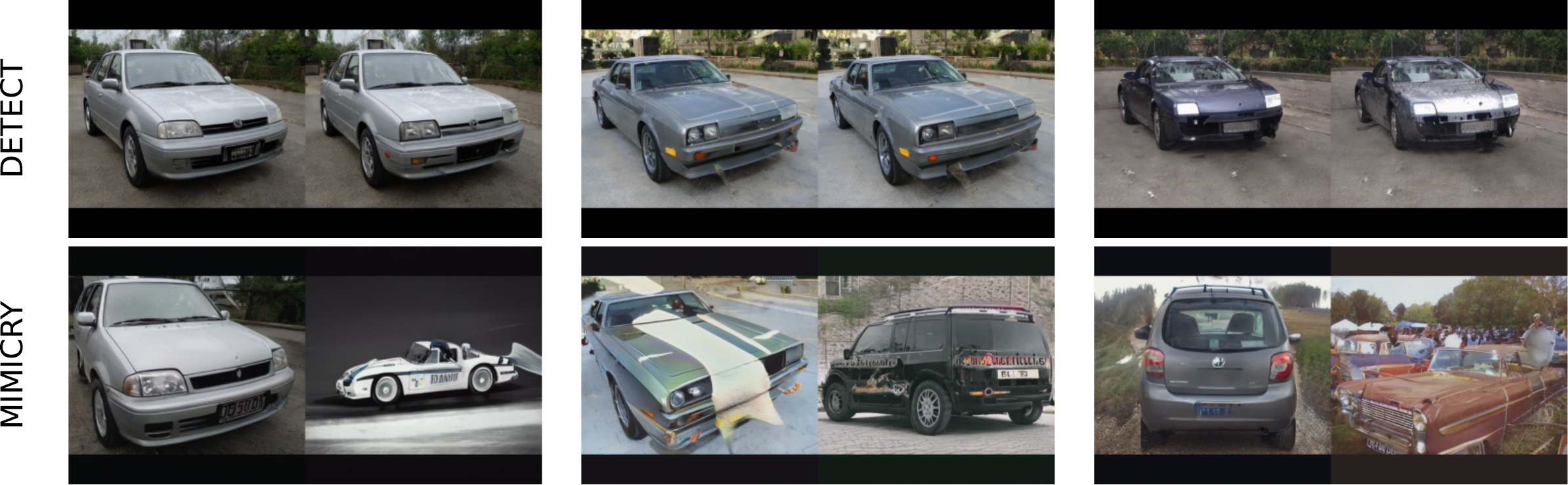}
        \caption{\cars Task: Make YOLO see anything, but a Car.}
        \label{fig:sub3}
    \end{subfigure}

    \begin{subfigure}[b]{0.7\columnwidth}
        \centering
        \includegraphics[width=0.95\columnwidth]{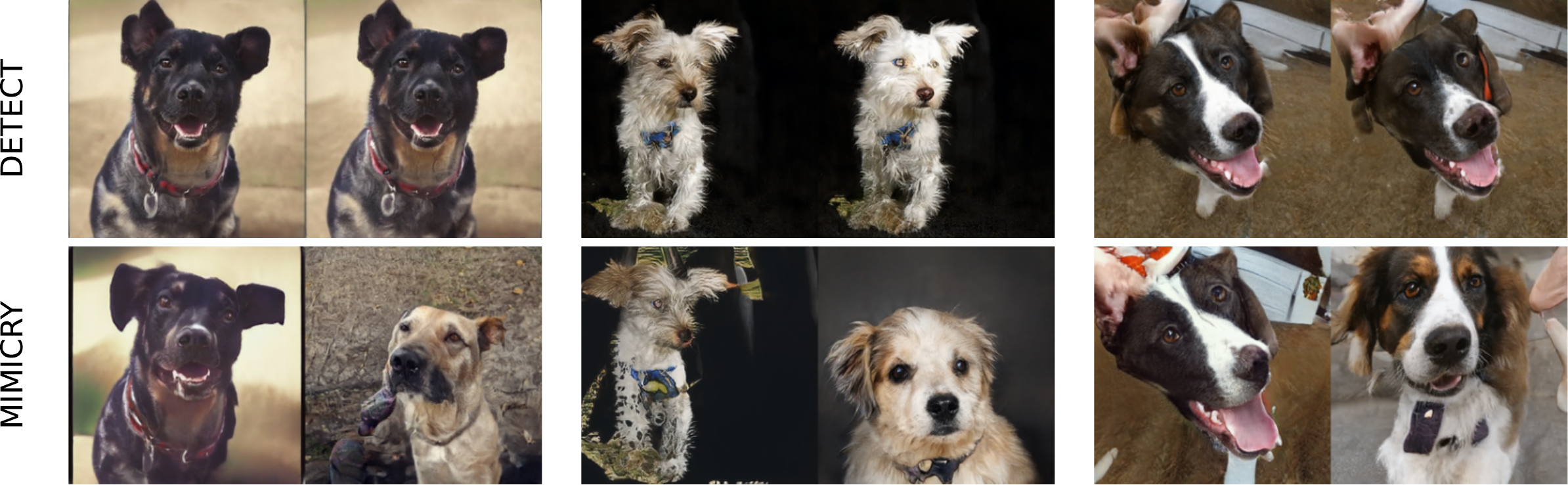}
        \caption{\dogs Task: Change the predicted Dog Breed.}
        \label{fig:sub4}
    \end{subfigure}
    \caption{Origin $\rightsquigarrow$ Perturbed Image Examples for \tool and \mimicry on each task.}  
    \label{fig:im_comp_all}
\end{figure}
\autoref{fig:im_comp_all} presents representative example pairs comparing
\tool and \mimicry. 
Across tasks, \tool produces more localized and attribute-specific perturbations, whereas \mimicry often alters multiple features and introduces additional unintended changes. In general, perturbations generated by \tool remain more controlled, modifying only the minimal region required for class transition.
These examples illustrate the qualitative differences underlying the quantitative trends reported in the main text.

\section{Full Prompt Examples Used for VLM Evaluation}
\label{appendix:prompts}

The following prompts were examples that used for the VLM-based relevance attribution and relabeling for eyeglasses detection task.\\

\head{Prompt for relevance attribution}
\begin{verbatim}

You are given three AI-generated images:

The first is the original face.
The second is generated by changing one latent direction.
The third is a difference mask showing where the change occurred.

Focus specifically on whether the presence of eyeglasses changes between 
the first and second images.
Outout a JSON object with the following keys:
    {"answer": "yes, eyeglasses are added/removed" 
       or "no, eyeglasses remain the same."}

Use the mask to support your reasoning and ignore irrelevant changes 
such as lighting or color tone.

\end{verbatim}

\head{Prompt for relabeling}

\begin{verbatim}
You are an image tagger for AI-generated portraits.
For each image, output a JSON object with the following fields: 
Determine whether the person in the image is wearing eyeglasses.
output a JSON object. 
{ "glasses": "Yes / No / Ambiguous"}
If the image quality is poor or unclear, respond with "Ambiguous".
\end{verbatim}



\end{document}